\documentclass[review]{elsarticle}

\journal{Journal of \LaTeX\ Templates}

\usepackage{graphicx}
\usepackage{caption}
\usepackage{subcaption}
\usepackage{color}
\usepackage{amsmath}
\usepackage{amssymb}
\usepackage{cleveref}
\usepackage{stmaryrd}

\begin{document}

\begin{frontmatter}
\title{Consistent time-to-failure tests and analyses of adhesive anchor systems}
\author[rvt]{Kre\v{s}imir Nin\v{c}evi\'{c}}
\ead{kresimir.nincevic@boku.ac.at}

\author[rvt]{Ioannis Boumakis}

\author[rvt]{Stefan Meissl}

\author[rvt,els]{Roman Wan-Wendner\corref{cor2}}

\ead{roman.wanwendner@ugent.be}

\cortext[cor2]{Corresponding author}
\address[rvt]{Christian Doppler Laboratory, University of Natural Resources and Life Sciences Vienna, Peter-Jordanstr. 82, 1190 Vienna, Austria.}
\address[els]{Magnel Laboratory for Concrete Research, Ghent University, Tech Lane Ghent Science Park - Campus A, Technologiepark-Zwijnaarde 904, 9052 Ghent, Belgium}

\begin{abstract}

Motivated by tunnel accidents in the recent past several investigations into the sustained load behavior of adhesive anchors have been initiated.
Nevertheless, the reliable life-time prediction of bonded anchor systems based on a relatively short period of testing still represents an unsolved challenge due to the complex non-liner viscoelastic behaviour of concrete and adhesives alike.

This contribution summarizes the results of a comprehensive experimental investigation and systematically carried out time-to-failure analysis performed on bonded anchors under sustained tensile load.
Two different adhesive materials that find widespread application in the building industry were used, one epoxy and one vinylester based.
Performed experiments include full material characterizations of concrete and the adhesives, bonded anchor pull-out tests at different loading rates, and time-to-failure sustained load tests. 
All anchor tests are performed in a confined configuration with close support.

After a thorough review of available experimental data and analysis methods in the literature the experimental data is presented with the main goals to (i) derive a set of recommendations for efficient time to failure tests, and (ii) to provide guidance for the analysis of load versus time-to-failure test data. 
Finally, a new approach based on a sigmoid function is proposed and compared to the established regression models. 
The analyses indicate a better agreement with the physics of the problem and, thus, more reliable extrapolations.

\end{abstract}


\end{frontmatter}


\section{Introduction}
\label{Intro}
Fastenings are crucial for the construction industry.
They facilitate the use of pre-manufactured elements in new structures and allow the attachment of load bearing structural components and equipment to existing structural elements. 
Typically, they are used in modular construction, structure renovations or retrofit projects to increase the ultimate strength or control deformations \cite{Eligenhausen_1,Cook_1,fib_retrofitting}.
Some applications require temporary solutions for a relatively short period of time (e.g. to restrain scaffolds).
Other problems require long-term solutions, and increasingly more often a 50-100 years service life time is demanded for infrastructure projects (e.g. ventilation systems, suspended ceiling panels,...).

Various types of fasteners can be found on the market that differ significantly in working mechanisms and/or installation time \cite{Eligenhausen_1}.
This contribution focuses on bonded anchors / adhesive anchors only. 
In general, bonded anchor systems are typically steel connectors such as threaded bars that are installed in holes, drilled or cored into a base material, and then filled with an adhesive.
These types of post-installed fasteners transfer loads mainly by adhesion / bond between the connection element (typically steel rebar, threaded bar, or anchor) and the base material \cite{Eligenhausen_1}.
In the recent past, two accidents \cite{Tunnel_accident, Sassaro} caused by the failure of adhesive fasteners under sustained load triggered a critical review of the respective approval and design procedures even though, ultimately, investigations showed that the main reason for both accidents was a wrong choice of adhesive in combination with bad installation. 

Obviously, limited ``short-time" testing is the only solution to satisfy modern building demands and to ensure the safe use of products on the market.
The engineering community has to rely on fundamental concepts and assumptions concerning the time-dependent material behavior, structural response and failure mechanisms since only short-term tests to investigate and ultimately predict the multi-decade behavior of building products are feasible.

Considering the complexity of the problem and the large number of influence factors including concrete aging, post-curing of the adhesives, non-constant boundary conditions, large variability in anchor types and geometries, it is obvious that reliable and accurate life-time predictions remain a quite challenging task.

Over the last few decades, different researches were performed and several standards have been established to predict the service life of adhesive anchors under sustained load \cite{ACI_Qualification,ETAG_MetalAnchors,ASTM_Standards,AC58}. 
The currently established approach is based on a pass/fail method using a displacement criterion applied to extrapolated relatively short sustained load tests \cite{Cook_2,Cook_3}.
The displacement criterion is derived from so-called confined tests, i.e. short-term pull-out tests with close support, and is defined as the displacement at maximum load.
The sustained load tests are performed in the same configuration by applying a percentage of the short-term ultimate capacity and by maintaining the load for a hold period of about 2,000 hours.
The obtained displacements for different load levels are then extrapolated in time to 50 years using either a logarithmic function \cite{ASTM_Standards,AC58}, or a power-law function \cite{ACI_Qualification,ETAG_001-5}.
Ocel et al. proposed even the use of a third degree polynomial function \cite{Ocel} without physical justification.

In general, this pass/fail method is assumed to provide a conservative time-to-failure prediction.
One reason is the conservative choice of the displacement criterion, i.e. the displacement at loss of adhesion (peak load) while the actual displacement at creep failure follows approximately the post-peak softening branch of short-term pull-out tests as shown by \cite{Muciaccia,DiLuzio}. 
Cook et al. \cite{Cook_2,Cook_3} and Davis \citep{Davis} showed that the main disadvantage of this method is the high dependence on the chosen creep function, and the inability to predict failure times for other anchor geometries or boundary conditions than the tested one.
An additional disadvantage reported by Wan-Wendner and Podrou\v zek \cite{RomanWW,Podrouzek} is the sensitivity of the regression to initial displacements and data point availability.

A rheological approach based on a modified Burger's model to predict the creep behavior of bonded anchor systems until failure was proposed by Kr\"ankel et al. \cite{TKrankel}.
This model considers both non-linear visco-elastic behavior and material degradation. It represents the anchor system as a single element without ability to redistribute stresses along the anchor rod and, thus, has to be calibrated on experimental data of the same diameter and embedment depth.

In an attempt to address the short-comings of the currently established approach Cook \citep{Cook_2,Cook_3, ACI_Cook_2017} and Davis \citep{Davis} proposed a method based on actual time-to-failure tests instead of trying to find experimentally sustained load levels that allow a displacement extrapolation to 50 years without violating the defined displacement criterion and, thus, approaching failure.
This approach requires a serious of sustained load tests at different relative load levels (with respect to the short term capacity), and the times to failure are monitored.
The main challenge lies in evaluating the obtained data due to (i) a significant and almost prohibitively large experimental scatter, and (ii) the limitation to relatively short failure times of typically much less than 1 year and the associated restriction to high relative sustained load levels. 

In the following Section~2 the commonly used regression models will be introduced. Section~3 presents the layout of the comprehensive experimental campaign followed by details concerning the material characterization in Section~4 and structural tests in Section~5. Finally, Section~6 utilizes the presented experimental data and present a systematic evaluation and comparison of all regression models concerning stability and predictive quality. The results clearly show the superior performance of sigmoid-shaped models that originate in their proximity to the physical nature of the problem.

\section{Regression models}
\label{Sec-Models}

For the analysis of time-to-failure test data Cook and Davis introduced a linear relationship between stress (relative load level) and logarithmic time to failure \citep{Cook_2,Cook_3,Davis,ACI_Cook_2017}. 
This corresponds to a logarithmic time to failure function used as formulated in Eq.~\ref{eq:CurrentApproach}.
        
        \begin{equation}
        	\label{eq:CurrentApproach}
            {y} = {a \cdot \ln(t_{f}) + b}
        \end{equation}
        
with $y = N/N_{100\%}$ the load level, $N$ the sustained load, $N_{100\%}$ the pull-out load, and  $t_{f}$ the failure time. 
Eq.~\ref{eq:CurrentApproach} uses natural logarithm of time which corresponds to an exponential decade of the load levels with parameter $a$ the half life of decade, and parameter $b$ the scale factor which also represents the theoretical load level at instantaneous time, $b=y(\lim_{t_{f} \to 0})$.
This conservative method works relatively well for high load levels, but can not accurately predict failure times for lower load levels, and exhibits some limitations.
Specifically, this logarithmic regression model violates the physical asymptotic properties for low and high loads and is solely able to represent data in the transitional domain. 
For very short load durations approaching $t=0$ the model predicts relative sustained load levels larger than 1.0, i.e. sustained load strengths that exceed the short-term strength \cite{IB_TertiaryCreep}. 
On the other hand, the logarithmic model will predict finite failure times also for very low load levels, and even for unloaded specimens since the regression line will intersect the abscissa.
This is not consistent with the established assumptions in concrete design that assumes linear creep, i.e. the absence of creep damage, up to approximately 40\% of the short-term strength/capacity \cite{IB_TertiaryCreep,Eurocode,ACI_209}.

In the literature a number of alternative regression models can be found that are either derived from theory or represent empirical models that are more or less infused by physical concepts. 

Very recently, the authors demonstrated that the logarithm of the creep rate of bonded anchor systems under sustained load is proportional to the logarithm of the failure time \cite{IB_CreepRate}. 
In combination with the power-law relationship between creep rate and relative applied stress level a power-law function for the relation of failure time on the load level is obtained. 
        
        
          \begin{equation}
        	\label{eq:S_PL}
            \ln(y) = \ln(b) + n\cdot \ln(t_{f})
        \end{equation}

        
where $y = N/N_{100\%}$, $b=y(\lim_{t_{f} \to 0})$, and $n$ is the slope in the log-log domain which translates to an exponent in linear plot. 
Coincidentally, this formulation (Eq.~\ref{eq:S_PL}) agrees with the power-law regression model proposed by Eligehausen et al. \cite{ACI SP283-9} which does not suffer from the more severe restriction of the logarithmic model, i.e. the regression model will always approach infinite failure time for an unloaded specimen. 
However, this linear regression model in logarithmic load versus logarithmic time domain still violates the upper asymptote.
Given the limitation to test data at relatively high load levels this shortcoming negatively influences the stability of the regression as it imposes an non-physical trend (linear in a log-log plot) onto the data that tend to approach asymptotically one for short failure times, i.e. exhibits initially a concave shape.

Boumakis et al. \cite{IB_TertiaryCreep} highlighted  the above mentioned disadvantages in an earlier work and proposed a potential solution based on time-to-failure experiments of concrete only (three point bending tests), see later discussion.
In this work they proposed a non-linear sigmoid function to establish a more appropriate relationship between stresses and failure times, as formulated in Eq.~\ref{eq:S_ShapeEq}: 

        \begin{equation}
        	\label{eq:S_ShapeEq}
            {y} = \kappa_{\infty} + (\kappa_{0}-\kappa_{\infty})\cdot\left(\frac{1}{1+b\cdot t_{f}}\right)^{c}
        \end{equation}

           
        where $y = N/N_{100\%}$ is sustained load level with respect to the short-term capacity, $\kappa_{\infty}$ defines the asymptotic level, $\kappa_{0}$=1, $b$ and $c$ are the fitting parameters, and $t_{f}$ is the failure time.
        
Additionally, Boumakis et al. \cite{IB_CreepRate} introduced a function based on the rate-theory. 
This can be formulated in the form of $N/N_{100}=f(t_{f})$ as
     
     \begin{equation}
        	\label{eq:S_RL}
            y = \kappa_{\infty}+\sinh^{-1}\left[\frac{(b\cdot t_f)^{n}}{c}\right]
        \end{equation}
where $y = N/N_{100\%}$, $\kappa_{\infty}$, $b$, $c$, and $n$ parameters to be fitted, and representing the asymptotic load level ($\kappa_{\infty}$), the onset of the asymptotic value ($b$ and $c$), and the slope of the decreasing load levels vs time ($n$).            
Furthermore, in this contribution an additional functional form is evaluated. 
As a mater of fact the Powell-Eyring function \cite{Powell_Eyring} is formulated in terms of load level versus failure time:
\begin{equation}
        	\label{eq:Eyring}
           y = \kappa_{\infty}+(\kappa_{0}-\kappa_{\infty})\frac{\sinh^{-1}(b\cdot t_{f})}{(b\cdot t_{f})}
        \end{equation}
where $y = N/N_{100\%}$, $\kappa_{\infty}$ the load level at which failure will occur in infinite time, $\kappa_{0}=1$ the load level at which failure will occur instantaneously, and $b$ a parameter to be fitted.

Despite the availability of several approaches and the effort of the research community, still more accurate models and methods are necessary  predict the behavior of fasteners under sustained load and develop an understanding concerning the main influence factors.
Additionally, a truly reliable prediction model should be able to account for the influence of anchor geometry (diameter, embedment depth), type of adhesive type, mechanical boundary conditions, and temperature.
These features are not yet available in any model that can be found in the scientific literature. 
A first step in this direction was recently made by Boumakis et al. \cite{IB_CreepRate} who successfully introduced a new type of analysis using a failure criterion that was previously established for materials such as alloys and polymers. 

After presenting a comprehensive experimental campaign on two adhesive anchor products this contribution aims at highlighting the most important influence factors affecting successful time to failure investigations and provides recommendations for testing and data analysis. 
After introducing the comprehensive experimental campaign performed by the authors and the corresponding results in Sections~3-5 the various regression models are finally systematically compared in Section~6.
Comparisons include the authors' data as well as data available in the literature.
The performed analyses clearly indicate that the proposed sigmoid model outperforms other regression models.
It satisfies the asymptotic physical behavior for high and low load levels while retaining a simple functional form.
Although this approach is still empirical the comparison with other established regression models for time-to-failure tests show improved stability and prediction quality.

\section{Experimental layout}
\label{Sec-ExpOverview}

A comprehensive experimental campaign was carried out comprising concrete characterization tests, adhesive mortar material tests, and structural tests performed on bonded anchors for two different chemical products. 
In total 50 concrete tests, 34 pull-out tests, and 50 anchor time to failure tests were carried out.

It is well know that concrete is an aging material, and that the material properties are improving with time due to ongoing hydration reactions, especially in the first few weeks.
The evolution of the concrete material properties are highly dependent on the temperature and humidity state inside the concrete member which in turn is influenced by the storage conditions.
The installation of the bonded anchors at significantly different times associated with different mechanical properties but also temperature and humidity states could affect the development of bond properties and certainly influences the short-term concrete capacity \cite{KN_Aging} and concrete creep contribution \cite{IB_SecondaryCreep,IB_TertiaryCreep} in sustained load experiments.

Due to the aforementioned reasons and limited availability of sustained load frames it was not possible to start all sustained load tests at the same time. 
Furthermore, it was desirable to minimize any aging effects while sustained load tests are ongoing.
Therefore, the bonded anchors were installed in $7$ months old concrete specimens.

All anchor tests were performed on cylindrical concrete specimens of sufficient size to prevent splitting (details are provided in later chapters) which had been stored inside the laboratory  ensuring a well-defined temperature and humidity state inside the concrete specimens. 
All anchor tests including preparation and installation were carried out in controlled environmental conditions with a constant temperature ($T$) of 22$\pm0.5^\circ$C and relative humidity ($RH$) of 50\% with a variations of $\pm3$\%,  

Structural tests on bonded anchors comprised both short-term pull-out tests in a confined configuration at different loading rates and long-term experiments. 
In the latter case the anchors were loaded hydraulically at a precisely controlled loading rate up to different relative load levels (with respect to the short-term pull-out capacity). 
After that the load was actively controlled to remain constant with a hysteresis of less than approximately 1~kN.
The displacement was continuously recorded at a sampling rate of 50~Hz for high load levels and tests expected to fail within hours, while for the lower load levels and longer tests the sampling rate was decreased but not lower than 1~Hz.

Two adhesive materials were investigated, further called Product~1 (epoxy based) and Product~2 (vinyl-ester based).
Only one anchor type, threaded bar M16, ($d_{nom}~=~16~$mm, bore hole diameter of 18~mm) and one embedment depth, $h_{ef}$, of 75~mm are considered in this experimental work.
The selected embedment depth corresponds to $4.7$ times the anchor diameter and also the maximum aggregate size.

\section{Material characterization}
\label{Sec-ExpMat}

Concrete and both adhesive mortars were fully characterized as part of this investigation and the larger project that this investigation is embedded in.
This section provides a summary of concrete and mortar properties in order to put the obtained structural tests into context.
Considering the extended test duration for time to failure tests the concrete mechanical properties were identified for different ages.
At 28 days only concrete cube compressive strength tests were carried out, while at 225 days (same age as sustained load tests with bonded anchors were started) concrete is fully characterized by means of unconfined compressive, tensile and fracture properties.
Additionally, creep and shrinkage tests were performed on concrete cylindrical specimens in parallel to the sustained load anchor tests with and without failure.
The creep and shrinkage tests include both sealed and drying shrinkage and creep tests in order to separate all four components that can be found in state of the art models \cite{Hubler2015797, Wendner2015771, wendner_statistical_2015, fib_code_2013}.

\subsection{Concrete mix design and curing kinetics}
\label{Sec-ConcMix}

For the purpose of this study a normal strength concrete was chosen with mix design parameters that represent a typical low-strength concrete used in the construction industry with a target strength class of C25/30. 
The used cement was a CEM~II~42.5~N.
The sieve curve was selected according to ETAG guidelines \cite{ETAG_AnnexA} and used a limestone coarse aggregate with a largest diameter of 16~mm.
The exact concrete mix design parameters are listed in Table~\ref{Table-MixDesign}.

\begin{table}[!h]
\centering
\caption{Concrete mix design parameters}
\label{Table-MixDesign}       
\begin{tabular}{llll}
\hline\noalign{\smallskip}
Mix design parameters  & Units & Concrete \\
\noalign{\smallskip}\hline\noalign{\smallskip}
coarse aggregate shape & [-] &  round \\
coarse aggregate type  & [-] & limestone \\
fine aggregates: 0-4 mm & [kg/m$^3$] & 1229.4  \\
coarse aggregates: 4-16 mm & [kg/m$^3$] & 828.9  \\
\noalign{\smallskip}\hline\noalign{\smallskip}
total amount of:  aggregates & [kg/m$^3$] & 2058.4   \\
~~~~~~~~~~~~~~~~~~~~~~~cement  & [kg/m$^3$] & 275.0  \\
~~~~~~~~~~~~~~~~~~~~~~~water   & [kg/m$^3$] & 166.3  \\
~~~~~~~~~~~~~~~~~~~~~~~superplasticizer & [kg/m$^3$] & 1.5  \\
~~~~~~~~~~~~~~~~~~~~~~~retarder & [kg/m$^3$] & 1.1  \\
\noalign{\smallskip}\hline\noalign{\smallskip}
 water/cement ratio & [-] & 0.60  \\
 aggregate/cement ratio & [-] & 7.5  \\
\noalign{\smallskip}\hline

\end{tabular}
\end{table}

\begin{figure*}[h!]
        \begin{center}
        \begin{tabular}{cc}
         \includegraphics[width=0.8\textwidth]{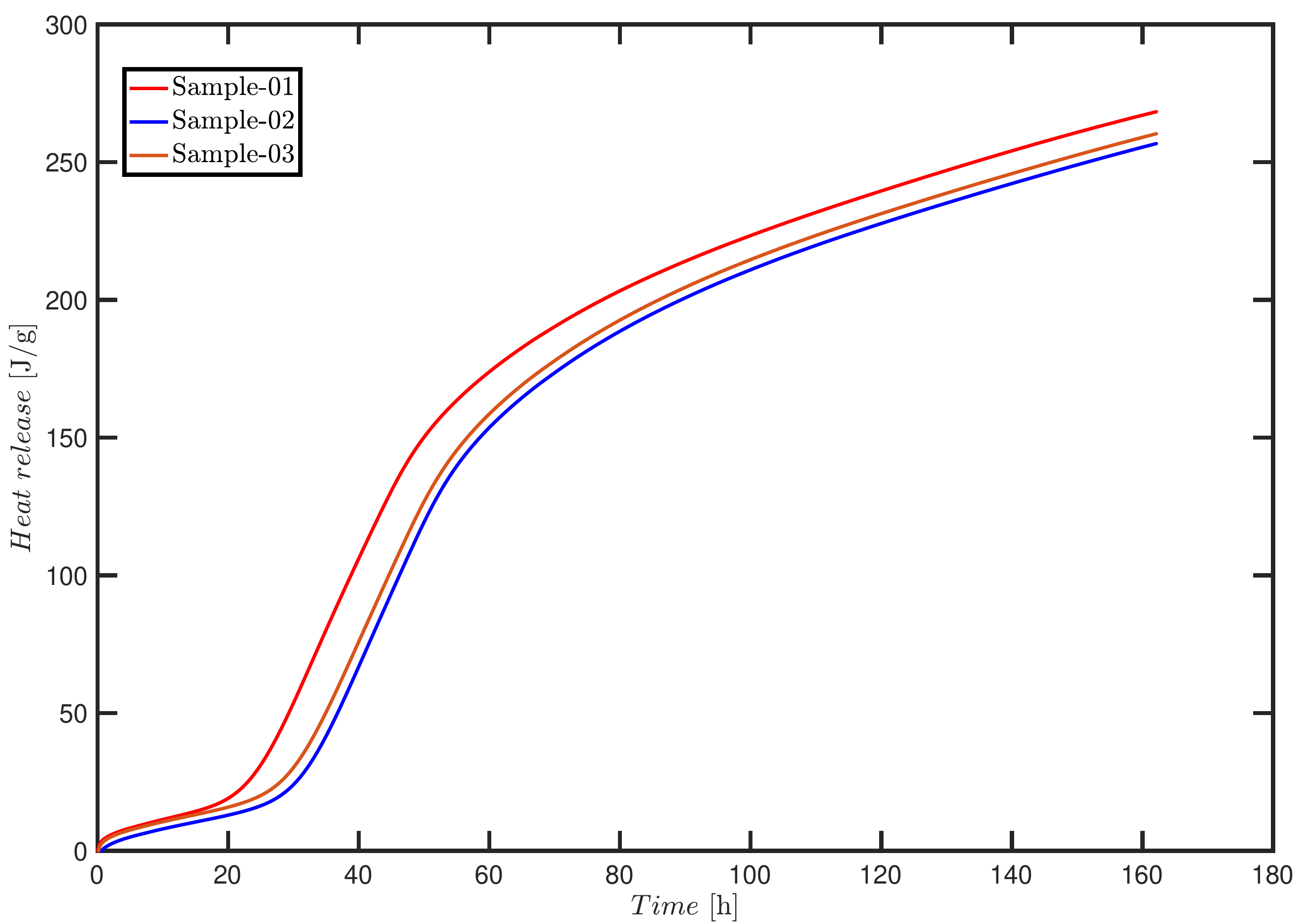}
 		\end{tabular}  
        \caption{Concrete calorimeter measurements}
         \label{fig:F1_Calorimetry}              
         	\end{center}          
        \end{figure*}

For the investigated concrete isothermal calorimetric measurements using a Tam~Air 8-channel calorimeter were performed in order to characterize the hydration properties of the concrete.
This gives essential insights into the aging behavior of the chosen concrete mix design, the development of mechanical properties and in particular the creep behavior.
Three tests were performed on the reactive constituents according to the mix design including water, binder, superplasticizer and retarder at a constant temperature of 20$^\circ$C. 
The heat release during the first 7 days was recorded every 30 sec.
Fig.~\ref{fig:F1_Calorimetry} shows the heat release results for all three repetitions revealing a quite consistent behavior.

\subsection{Concrete curing and storage}
\label{Sec-ConcCure}

After casting, all concrete specimens were kept in the moulds for 24 hours.
Later, the concrete specimens used for the material characterization (cubes, cylinders and prisms) were stored in lime-saturated water until the tests were performed \citep{BSI,ASTM_Curing} at ambient lab temperature of approximately 20-25$^\circ$C.
The cylindrical concrete specimens used for the structural tests were dry cured under laboratory conditions, according \cite{ETAG_AnnexA}.

\subsection{Concrete mechanical properties}
\label{Sec-ConcMech}

Concrete mechanical properties were obtained at 28 and 225 days.
At the age of 28 days only compressive strength ($f_{c,150}$), measured on cubes with a side length of 150~mm, was determined according to \cite{BSI_Compressive}.
At the second age, concrete was fully characterized in terms of compressive strength ($f_{c}$) and modulus ($E$), splitting tensile strength ($f_{t,spl}$), and total fracture energy ($G_{F}$).
Contrary to the 28 days tests, for the later age compressive strength was obtained both from standard cubes ($f_{c,150}$) and cylinders ($f_{c,cyl}$) with 150~mm diameter and 300~mm height according to \cite{BSI_Compressive}.
Brazilian splitting tests \cite{BSI_Tensile} were performed to obtain the indirect tensile strength ($f_{t,spl}$) on cylindrical specimens with a diameter of 150~mm and a height of 70~mm.

The total fracture energy was obtained from three point bending tests controlled by crack mouth opening displacement (CMOD) using an extensometer of type Epsilon 3542-050M-025-HT2.
Notched prisms with dimensions of $100 \times 100 \times 400$~mm, a relative notch depth equal to 30~mm, notch width of approx. 4~mm, and a span of 300~mm were tested with a constant opening rate equal to 0.0001~mm/s.
The displacement field of the specimen was recorded by means of 3D digital image correlation (DIC).
The fracture energy was determined following the work of fracture method, using the displacements measured by DIC.
The relative load point displacement is determined as difference between displacement field under the load point and the two displacement fields above the supports, as explained in detail in \cite{LC_SLA}.
 
All experimentally obtained mechanical properties are listed in Table~\ref{Table-Concrete Properties} together with the corresponding coefficients of variation (CoV).
Each mean value represent a mean value of 4-5 tests.

\begin{table}[h!]
\caption{Experimentally obtained material properties}
\label{Table-Concrete Properties}       
\begin{tabular}{lllll}
\hline\noalign{\smallskip}
Age [d] & $f_{c,150}$ ~[MPa] & $f_{c,cyl}$ ~[MPa] & $f_{t,spl}$ ~[MPa] & $E_{cyl}$ ~[GPa]  \\
\noalign{\smallskip}\hline\noalign{\smallskip}
28 & $39.8 \pm 6.4 \%$ & - & - & -  \\

\noalign{\smallskip}\hline
225 & $53.7 \pm 6.3 \%$ & $49.83 \pm 5.9 \%$ & $4.7 \pm 3.2 \%$ & $49.8 \pm 5.9 \%$ \\

\noalign{\smallskip}\hline
\end{tabular}
\end{table}

\subsection{Concrete creep and shrinkage}
\label{Sec-ConcCS}

Apart from the standard concrete characterization tests the experimental program also included creep and shrinkage tests for the investigated concrete.
For that purpose, concrete cylindrical specimens with a diameter of 150~mm and a height of 300~mm were cast. 
After being removed from the moulds they were kept together with the specimens for the standard characterization in lime saturated water until the age of loading (225 days).
Each specimen was instrumented equally with three 50~mm long strain gauges distributed uniformly around the perimeter of the cylinder and aligned with the axis of the cylinder at the midsection to measure the creep and shrinkage strains.
All creep and shrinkage tests were performed in an environmentally controlled room together with the sustained load anchor specimens at a constant temperature $T=$22$\pm0.5^\circ$C and relative humidity $RH=50\%\pm3\%$.

Creep tests were carried out for two sealed and two drying specimens.
All specimens were loaded hydraulically in a creep frame up to 30\% of the cylinder compressive strength ($f_{c,cyl}$) obtained on companion specimens at the same day at an age of 225~days.
Shrinkage tests were performed on two drying and two sealed specimens.
The two specimens used for autogeneous shrinkage were instrumented just after being demoulded 24 hours after casting.
All concrete cylindrical specimens used for creep and shrinkage tests had the same geometry and were instrumented in the same way with strain-gauges.
More detailed information about creep and shrinkage tests and drawings of the creep frame can be found in \cite{IB_SecondaryCreep}.

\begin{figure*}[h!]
   	   \begin{center}
    	  \begin{tabular}{cc}    
     \includegraphics[width=0.45\textwidth]{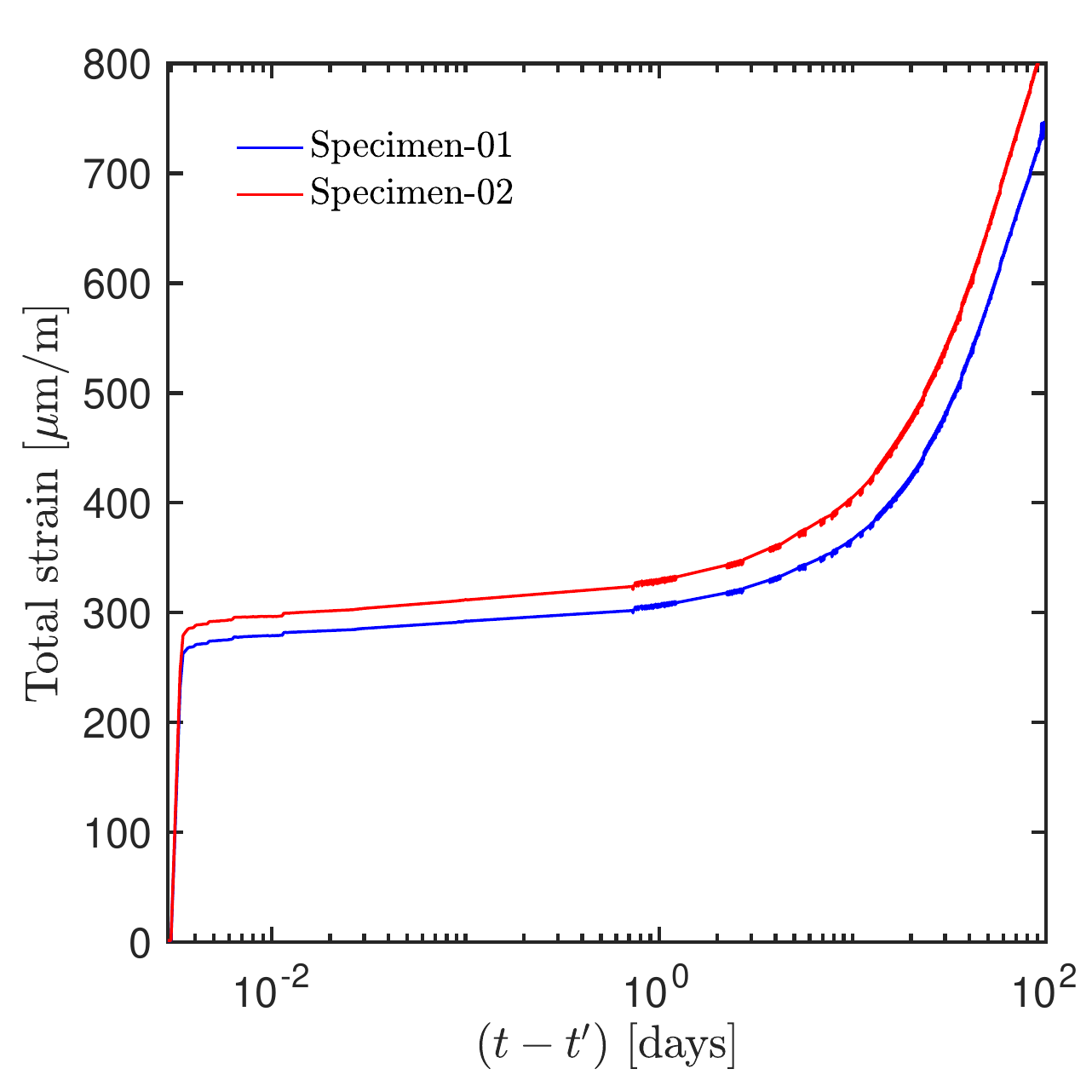} &
     \includegraphics[width=0.45\textwidth]{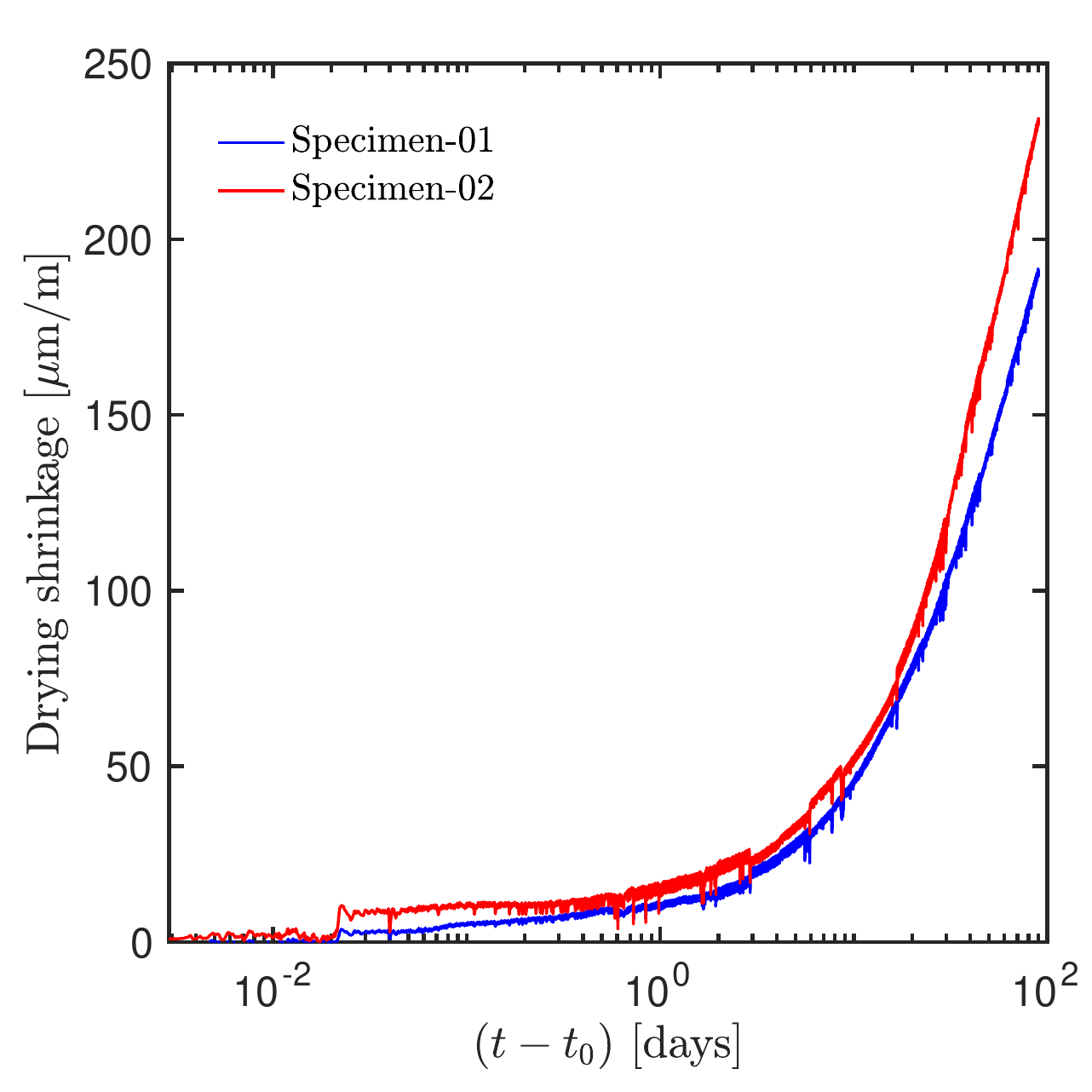}\\
         (a) & (b) \\
		\end{tabular} 
       \caption{Experimentally obtained a) drying creep and b) drying shrinkage measurements}
        \label{fig:CreepAndSh}       
      	\end{center}          
      \end{figure*}

Fig.~\ref{fig:CreepAndSh} shows the total strains obtained from a) loaded drying specimens (creep), and b) unloaded drying specimens (shrinkage). Each curve is the average strain time history of three strain gauges attached to a concrete specimen.


\subsection{Concrete rate-dependent damage and time-to-failure}
\label{Sec-ConcTTF}

Finally the time-dependent fracture properties of concrete have been tested at an age of 223 days. 
In particular, notched specimens with the same geometry as those used in the fracture energy tests (see Section \ref{Sec-ConcCS}) were used to determine the loading rate effect on strength. 
For the rate tests the prismatic specimens were loaded in CMOD control ten faster and ten time slower than in the standard fracture energy tests. 
The increase of the loading rate resulted in an increase of the peak force, which was expected according to previous studies, \cite{Watstein, Hughes, Yon, reinhardt_1985, Mainstone1975, Dilger, Bischoff1991}. 
Additionally, also sustained load tests with failure were performed.
In these tests each of the prisms was loaded to a target value, then the load was kept constant until failure of the specimen was observed. 
Finally the failure times were recorded. In total $14$~prisms were tested. 
The sustained load levels covered the range between $82\%$ and $\sim 100\%$ with increments of $2.5\%$.
The observed failure times found to be in a range of $18.17-149,839$~sec.

\subsection{Adhesive properties}
\label{Sec-Adhesives}

In this comprehensive experimental program two adhesives, representative for the construction market, were used for the anchor tests but also fully characterized as materials.
The later is essential to develop a thorough understanding of the relevant mechanisms and to develop reliable numerical models.
Product~1 (later P1) is an epoxy-based mortar based on a bisphenol-A/F epoxy resin with an amine hardener, while Product~2 (P2) is vinyl-ester based  with a dibenzoyl peroxide hardener.
Both products contain inorganic filler materials (quartz, cement) with a relatively high amount (20-50 vol\%).

The still ongoing experimental characterization includes the short-term elastic and fracture-mechanical properties as well as long-term (creep, and time to failure) tests for both products.
The goal of this investigation is not only to characterize the materials at one age, but also to provide insights into the time-dependent changes of mechanical properties for different curing conditions. 

Mechanical properties of two used adhesive materials were characterized by Singer et al. \cite{GSinger_Aging} considering different storage time and post-curing temperatures.
The obtained material properties are listed in Table~\ref{Table-AdhesiveProperties} for both investigated products.

\begin{table}[h!]
\centering
\caption{Mechanical properties experimentally obtained for the two adhesive products post-cured at room temperature (23$^\circ$) for 24h}
\label{Table-AdhesiveProperties}       
\begin{tabular}{llll}
\hline\noalign{\smallskip}
Product &  Young's modulus & Tensile strength & Poisson's ratio  \\
        &  [MPa]           & [MPa]            & [-]              \\
\noalign{\smallskip}\hline\noalign{\smallskip}
P1 & 6816.7 & 64.2 & 0.36  \\
P2 & 5580.2 & 13.9 & 0.33  \\
\noalign{\smallskip}\hline\noalign{\smallskip}
\end{tabular}
\end{table}

All experimentally observed mechanical properties for other temperature cases are in detail reported in \cite{GSinger_Aging}, showing the substantial changes of properties in course of time while the curing related volume changes are reported in \cite{GSinger_Shrinkage}.
Mode~I fracture properties are reported in Marcon et al. \cite{Marco_Mode1}.

Additionally, 50 year master creep curves have been constructed for tensile and shear modulus using the time-temperature shift methodology as reported in \cite{Linz_MCCurve,Linz_ShearTest}, for both investigated mortars albeit only at a fully cured state.

\section{Bonded anchor tests}
\label{Sec-BA}

Structural pull-out tests were performed for both previously introduced adhesive mortar products.
As part of this comprehensive investigation pull-out test in a so-called ``confined" and ``unconfined" configuration were performed.
The latter results in concrete cone failure.
Although this data is available for the investigated concrete batch and both anchor products it is not the focus of this investigation and, thus, it is not further discussed. 
In this contribution all reported pull-out results are tested in a ``confined configuration" with supports close to the anchor.
Almost all short and long-term tests of this campaign were performed on cylindrical concrete specimens, one anchor per specimen.
For consistency reasons the short-term pull-out capacity was determined on the same specimen geometry that later is used for the sustained load tests. 
Additionally, standard confined tests in a concrete slab were also performed in closed-loop displacement control.
The loading rate influence on bonded anchor tests was determined both in concrete slabs for the ``confined" and the ``unconfined configuration".

\subsection{Anchor installation}
\label{Installation}

This type of experiments and measurements are very sensitive and highly dependent on the boundary conditions.
It is well know that the cleaning and moisture of the hole during the installation can significantly affect the anchors' performance \cite{CookAndKonz}.
Thus, in this study all known and relevant influence factors were carefully taken into account to minimize the scatter as much as possible.
Therefore, concrete specimens (geometrical details provided in \ref{Anchors}) were stored inside the laboratory to ensure the same temperature history with a minimum of variations for all specimens.
The anchors were installed at a concrete age of 7 months and were loaded consistently always 24 hours after the installation.
All the installation steps regarding the preparation, drilling and cleaning hole were carefully carried out following the product specific guidelines by the same two researchers in concrete cylinders after reaching thermal equilibrium.
Even though high quality drill bits with ``PGM" certificate were used for drilling the holes, always a new drill bit is used for different type of tests (e.g. sustained load, or loading rate sensitivity tests).
The installation and testing procedure were carried out in the temperature controlled room with a constant temperature ($T$) of 23$\pm2^\circ$C and relative humidity ($RH$) of 50\% with a variations of $\pm3$\%.

\subsection{Short-term tests and loading rate sensitivity}
\label{Peaks}

Apart from the concrete curing history, and possible imperfections due to the installation procedure, a crucial influence factor is the loading rate sensitivity of the short-term pull-out capacity \cite{Ozbolt_LoadingRates}.
Both, concrete and adhesive polymers are rate-dependent materials \cite{IB_TertiaryCreep, Rate_Polymers_1,Rate_Polymers_2}.
Thus, it is crucial to control the loading rate in all applications. 
Prior to the sustained load test investigation, a systematic loading rate effect study was performed to quantify the effect of the loading rate on the pull-out capacity.
All tests were performed for both adhesive products in confined and unconfined configurations using concrete slabs with dimensions of 150x100x30~cm cast in the same batch as the specimens used for the material characterization and the cylindrical slabs for the time to failure tests.
Tests were carried out using the testing load frame with a 630~kN servo-hydraulic jack. 
The tensile load was applied at the top of the anchor, controlled by one of two linear variable differential transformers (LVDT: HBM IWA/20 mm-T) with a nominal measuring range of 20 mm.
Two LVDT's were installed close to the steel confinement plate in case of ``confined" tests, and close to the concrete surface in case on ``unconfined" tests to ensure stable post-peak response and a constant displacement rate.
For this purpose, bonded anchors with the same geometry and embedment depth were used as tested under sustained load.
Since this study is focused on the sustained load behaviour of bonded anchors in a confined configuration, only the related loading rate effect results are discussed here.
Apart from the so-called ``quasi-static" rate (0.008~mm/s) that according to approvals has to reach the peak load between 1-3 min, another two loading rates were considered: (i) ``low rate" with the speed of 0.0008~mm/s, and (ii) ``high rate" with the speed of 0.08~mm/s.
The result of the obtained pull-out capacities with regard to the loading rate are listed in Table~\ref{table-RateEffect} with the corresponding coefficient of variation (CoV).
Each value is the average of four tests.

\begin{table}[h!]
\centering
\caption{Result of the loading rate effect on pull-out capacity in confined configuration}
\label{table-RateEffect}       
\begin{tabular}{llll}
\hline\noalign{\smallskip}
Product  & Low rate & Quasi-static rate & High rate   \\
\noalign{\smallskip}\hline\noalign{\smallskip}
Product 1  & 117.4 kN $\pm$ 2.6\% & 129.3 kN $\pm$ 5.1\% & 146.6 kN $\pm$ 1.9\% \\
Product 2  &  78.9 kN $\pm$ 6.7\% &  85.8 kN $\pm$ 4.4\% &  89.3 kN $\pm$ 6.0\% \\
\noalign{\smallskip}\hline\noalign{\smallskip}
Time to peak  & 27 min & 2.4~min & 0.3~min  \\
\noalign{\smallskip}\hline\noalign{\smallskip}
\end{tabular}
\end{table}

As it can be seen from Table~\ref{table-RateEffect}, the results indicate the presence of a statistically significant loading rate effect.
Based on linear regression in a plot of peak load versus logarithmic loading rate the loading rate sensitivity can be quantified. 
An increase in loading rate by a factor of 10 causes an increase in pull-out capacity by $4.8\%$ and $3.1\%$ for Products~1 and 2, respectively.
It has to be noted that the rate effect may follow another functional form if a wider loading rate range is taken into account.

        

Typically, standard confined pull-out tests to determine the short-term capacity are performed either in load or displacement control. 
However, not much attention is placed on the actual load rate as the effect on the peak load in the range of ``normal" loading rates is of the same order as typical experimental scatter.
For sustained load tests (with and without failure) a completely different experimental set up is used in which the load is applied by pressurized air, hydraulics or even torque in combination with spring systems.
Most of the reported literature data sets on sustained load tests are carried out using hydraulic jacks or compression springs to sustain the load, often in combination with hand pumps \citep{Cook_2}.
Even if automatic hydraulic loading systems are available they typically can not be operated in displacement control.
As a result, the loading rate is generally neither controlled, nor consistent with the rate used for the determination of the short-term capacity.

Consequently, the resulting inconsistencies between the loading rates used for the determination of the short-term capacity (100\%) and later for sustained load (long-term) tests will lead to inaccurate definitions of relative load levels with respect to the short term tests.
Clearly, as a result the observed failure times for a given load level would be unreliable.
Considering the intrinsically large experimental scatter, that is typically observed in time to failure tests, the avoidance of any such additional and especially systematic error is detrimental for a successful experimental campaign.
Therefore, in this study the same loading rate was used both for the peak identification and sustained load tests.
Furthermore, the short-term capacities (100\% load level, $N_{100\%}$) were not only determined in slabs but also the same specimen geometry that is later used for the time to failure tests.

Based on the quasi-static short-term results, the load levels for the sustained load tests were selected, taking into account the observed scatter and potential probability of failure during loading for the relatively high load levels.
Table~\ref{table-Pull-out peaks} summarizes the experimentally obtained mean peak values for both products with the corresponding CoV.
Each peak value represents a mean value of four and six tests for Product~1 and Product~2, respectively. 

\begin{table}[h!]
\centering
\caption{Experimentally obtained pull-out load capacities for both products}
\label{table-Pull-out peaks}       
\begin{tabular}{llll}
\hline\noalign{\smallskip}
Product & Tests & $N_{100\%}$ [kN] & CoV [\%]   \\
\noalign{\smallskip}\hline\noalign{\smallskip}
Product 1 & 4 & 157.3 & 3.36\% \\
Product 2 & 6 & 111.8 & 7.48\% \\
\noalign{\smallskip}\hline\noalign{\smallskip}
\end{tabular}
\end{table}

As it can be seen from Table~\ref{table-Pull-out peaks}, for the same embedment depth ($h_{ef}$) and anchor size (M16) tests for Product~1 resulted in a higher pull-out capacity and lower scatter in terms of the coefficient of variation (CoV).
Based on the uniform bond model \cite{Cook_UniformModel}, the bond strength of Product~1 is $\tau_{1}~=~31.3$ MPa, while for Product~2 $\tau_{2}~=~22.2$~MPa is obtained.

\subsection{Sustained load tests}
\label{Anchors}

\begin{figure*}[h!]
        \begin{center}
        \begin{tabular}{cc}
         \includegraphics[width=0.8\textwidth]{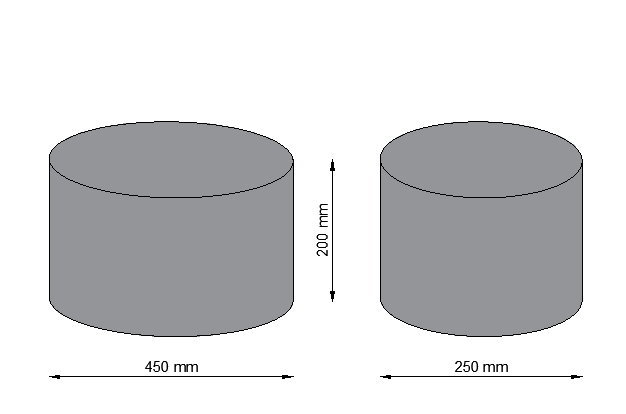}
 		\end{tabular}  
        \caption{Geometry of concrete specimens used for a) Product~1, and b) Product~2}
         \label{fig:SpecimenSizes}              
         	\end{center}          
        \end{figure*}

Sustained load experiments were carried out on non-reinforced cylindrical concrete specimens.
Based on the experience gained in the previous comprehensive experimental campaign, different sizes were used for the two products.
The aim was to avoid potential splitting failure due to the evolution of lateral stresses, while optimizing the weight and ensuring easy manipulation.
Finally, two different cylindrical concrete specimens were used: (i) specimens with 450~mm diameter and 200~mm height for Product~1, and (ii) specimens with 300~mm diameter and 200~mm height for Product~2.
Fig.~\ref{fig:SpecimenSizes} shows sketches of the selected specimen geometries for each product.

\begin{figure*}[h!]
        \begin{center}
        \begin{tabular}{cc}
         \includegraphics[width=0.85\textwidth]{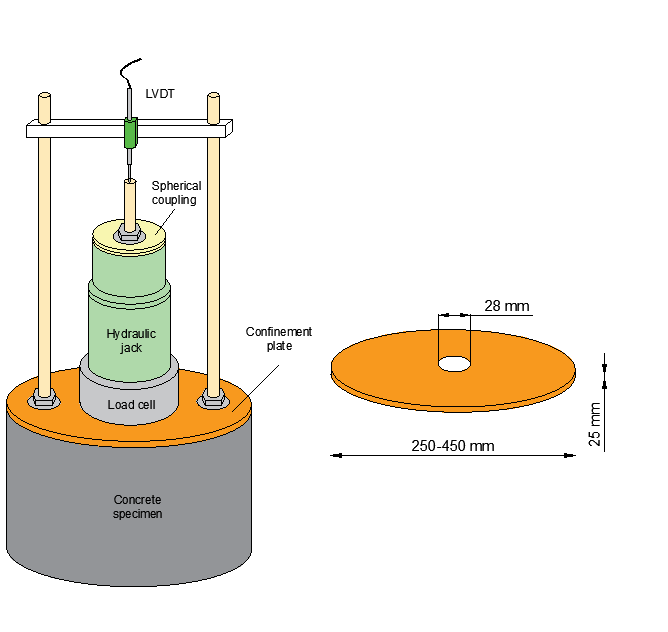}
 		\end{tabular}  
        \caption{Experimental set-up used for sustained load tests}
         \label{fig:Setup_sketch}              
         	\end{center}          
        \end{figure*}

In this experimental investigation pull-out tests on post-installed bonded anchors in a so called ``confined" configuration were carried out \citep{EAD_330499,EOTA_TR048}.
Contrary to the standard ``unconfined" configuration \cite{EAD_330499,EOTA_TR048} where the formation of a concrete cone is almost unrestricted, the confined configuration avoids concrete cone failure and ensures almost exclusive bond failure.
Typically, ensuring a sufficiently high steel strength, this leads to the adhesive bond failure localized at the inner (failure on the interface between steel anchor rod and adhesive layer), outer (failure on the interface between adhesive layer and concrete), or a combination of the two failure modes.
Fig.~\ref{fig:Setup_sketch} shows the detailed sketch of the set-up used for sustained load tests.

The load was applied using the hydraulic system of 700~bar capacity with 12 individually actively controlled jacks with a load capacity equal to 22~t.
Each hydraulic jack is equipped with a spherical coupling in order to minimize the bending moments due to installation and set-up imperfections.
Confinement plates of 25~mm thickness with an outer diameters of 250 and 450~mm (the same as concrete cylinders), respectively, were used to ensure a confined configuration following the anchor related codes \citep{EAD_330499,EOTA_TR048}.
The inner diameter of the plate was 28~mm that corresponds to 1.75 times anchor diameter ($1.75 \cdot d_{nom}$).
In order to measure the actual system response and creep deformations after the load is applied, the displacement was measured on the top of the anchor relative to the steel confinement plate using linear potentiometers of type LRW2-C-10 with a maximum range of 10~mm and the linearity of $\pm0.3$\%.
The linear potentiometer was carefully set ensuring alignment with the axis of the anchor.
For the short-term tests calibrated load cells were used in addition to the hydraulic pressure sensors while for the long-term tests on the pressure development was monitored.
The corresponding experimental setup for the sustained load tests corresponds to the one shown in Fig.~\ref{fig:Setup_sketch} but without the load cell.

\subsection{Definition of load levels}
\label{load levels}

Based on the observed short-term pull-out capacities ($N_{100\%}$) and corresponding coefficients of variation (CoV), different load levels were selected for the two products.
For Product~1, the load levels used to perform sustained load tests were: 95\%, 85\%, 75\%, 65\%, and 60\% with respect to the corresponding $N_{100\%}$.
Since the failure times for Product~1 at moderate load levels exhibited comparably low scatter, tests at the relatively high load level of 95\% were added.
For Product~2 time to failure tests were performed at 85\%, 80\%, 75\%, 70\%, and 65\% of the short-term capacity.
Due to the comparably larger scatter for this product it was not possible to test at higher load levels.

\subsection{Time-displacement curves}
\label{LD curves}

\begin{figure*}[h!]
        \begin{center}
        \begin{tabular}{cc}    
       \includegraphics[width=0.45\textwidth]{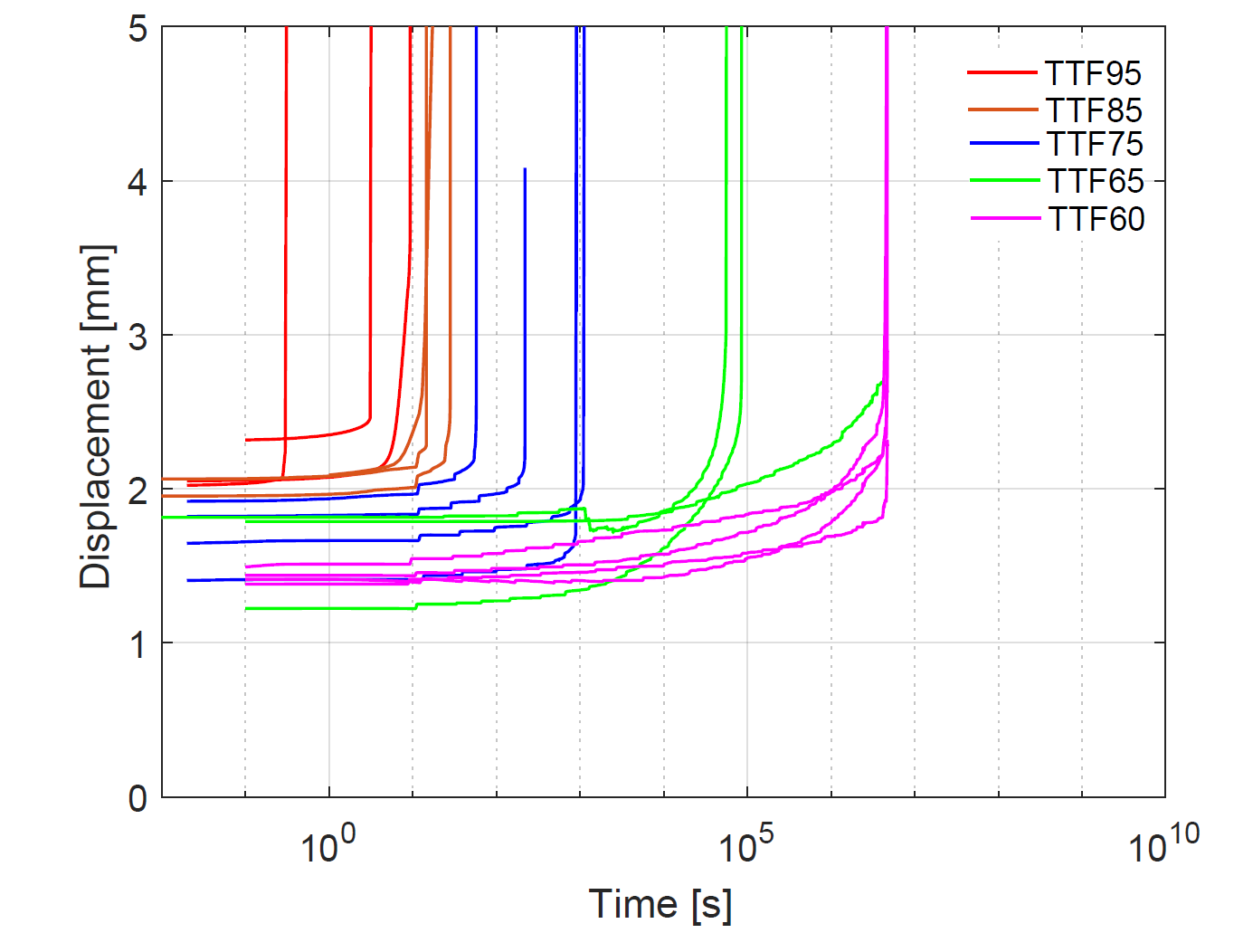} &
       \includegraphics[width=0.45\textwidth]{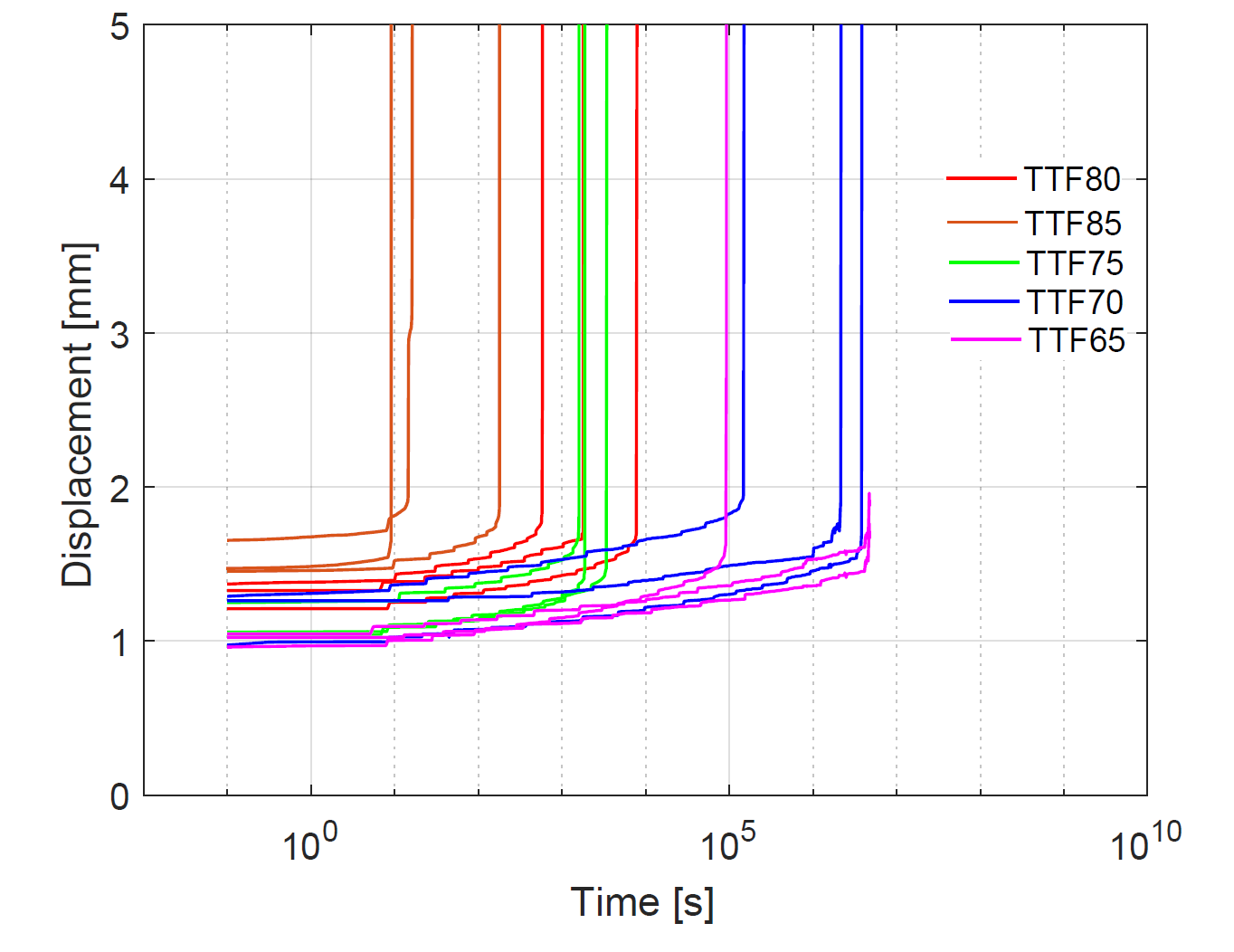}\\
          (a) & (b) \\
		\end{tabular} 
        \caption{Time-displacement curves for a) Product 1, and b) Product 2}
        \label{fig:LD_Curves}       
         	\end{center}          
        \end{figure*}

In the following chapter typical time-displacement curves are shown for both products and all load levels. 
It is well know that the deformations in a bonded anchor system subjected to a constant load can will increase with time due to the visco-elastic nature of both concrete and adhesive mortars. 
At higher load levels of approximately more than 30-40\% the development of damage will lead to an over-proportional increase in creep deformations with increasing load (nonlinear creep). 
As soon as cracks are formed they propagate, leading to further stress redistribution, growing deformations, and ultimately progressive collapse.
Clearly, the evolution of deformation rate and the failure time highly depend on the stress state.
Typically, three different stages of creep can be identified: primary, secondary and tertiary creep \cite{TKrankel}.
The first is characterized by quickly growing deformations, the second by a stable creep process with an almost constant creep rate while the third one is defined by progressively growing deformations that lead to failure, see Boumakis et al. for a more detailed discussion \cite{IB_TertiaryCreep}.
Fig.~\ref{fig:LD_Curves} shows typical time-displacement curves for both products, with time=0 after the load was fully applied.

Both figures clearly show different creep stages, where the duration and slope of the secondary creep phase is dependent on the load level/stress state.
Fig.~\ref{fig:LD_Curves}  clearly illustrates the expected trend. 
The duration of the secondary creep stage is related to the load level: the higher the load level, the shorter the secondary creep stage, and therefore the failure time. 
The trend among the curves is more apparent for Product~1 (Fig.~\ref{fig:LD_Curves}a), while the bands of experimental curves for Product~2 (Fig.~\ref{fig:LD_Curves}b) partly overlap. 
This can be explained by the higher scatter observed in the short-term tests of Product~2 which is (i) an indication of the larger inherent uncertainties, and (ii) is also propagated into the load level definition.

\subsection{Definition of loading time}
\label{loading time definition}

In sustained load tests the load is applied more or less rapidly but always over a finite period of time. 
As soon as the first load increment is applied creep deformations can be observed.
For the data analysis one can assume either a step function corresponding to the instantaneous application of the full sustained load, a loading ramp of constant rate, or the actual stress history.
In the latter two cases the solution of the Volterra integral equation, i.e. superposition of creep histories for the individual stress increments, is required while the assumption of a step function allows a direct calculation of creep deformations based on a compliance function.

Furthermore, it is not trivial to determine the time of full load application, i.e. the origin of the time axis for the data analysis and determination of failure times as this is a gradual process. 
Generally, one could assume the midpoint of the loading ramp or the time of full load application.
Differences between the approaches and with the observed behavior become more apparent when the load is applied more slowly. 

They are especially important for high load levels for which the failure times are in the same order of magnitude as the duration of load application but lose importance for low load levels associated with high failure times.
The previously discussed choices may lead to significant differences in failure times, reflected in a horizontal shift of data points, that primarily affect tests at high load levels. 
Considering also the logarithmic nature of the failure times, this horizontal time shift has a pronounced effect on the time to failure plot, and ultimately on the fitted stress versus time to failure curve.
For  consistency reasons, in this contribution a rather quick but constant loading rate of approximately 12-15~s was chosen and used for all tests.
All reported failure times are determined with regard to the time of full load application. 
This time is taken as the instant when the load applied by the hydraulic jack reaches the target value for the first time. 

\subsection{Time to failure determination}
\label{TTF times}

\begin{figure*}[h!]
        \begin{center}
        \begin{tabular}{cc}
         \includegraphics[width=0.75\textwidth]{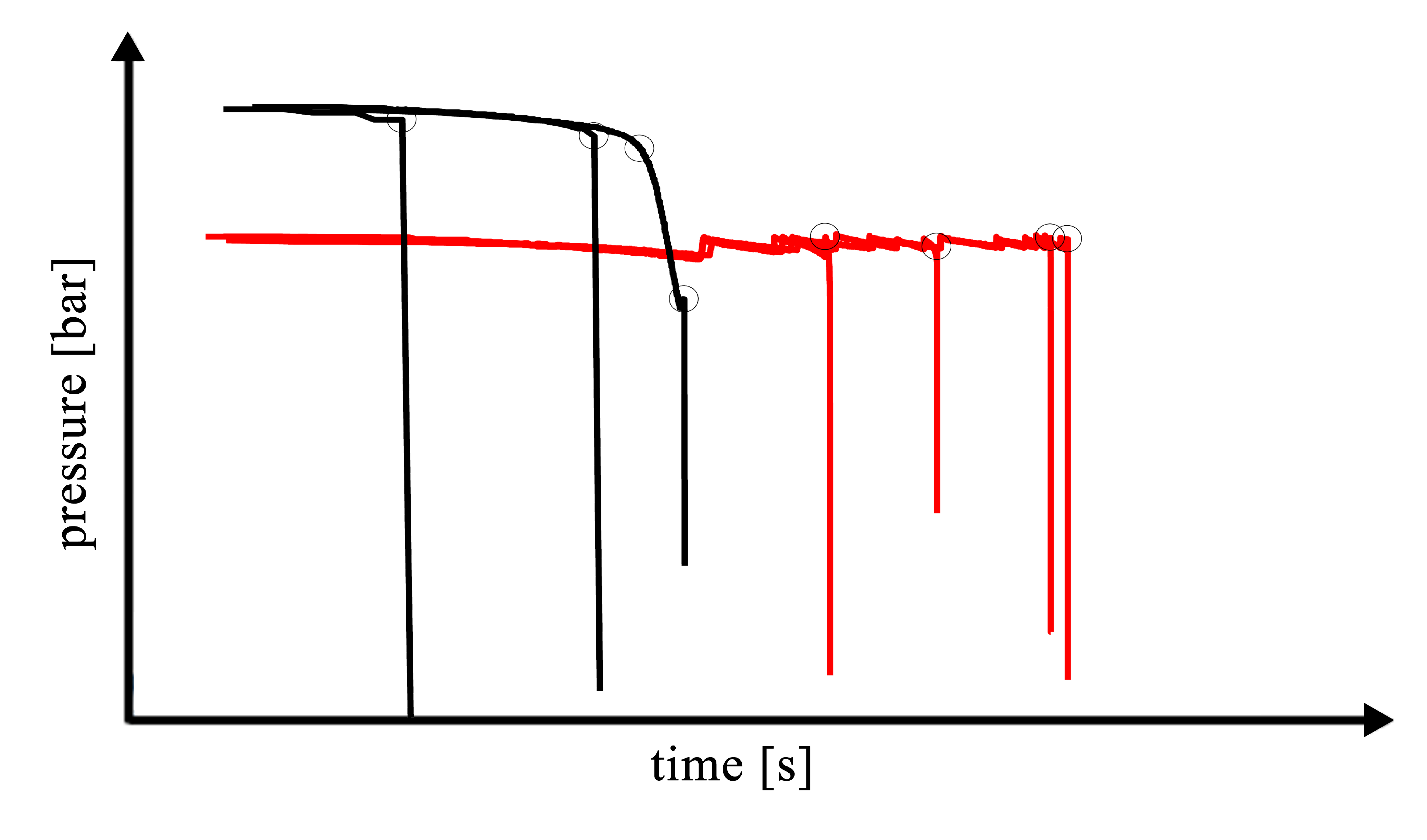}
 		\end{tabular}  
        \caption{Schematic view of pressure drop in time used to determine time to failure}
         \label{fig:PressureDrop_Sketch}              
        	\end{center}          
     \end{figure*}

Prior to analyzing stress versus time to failure data it is important to define exact failure times.
These are determined by the time of full load application as discussed in the previous chapter but also the time at which as specimen is considered to be failed.
In this contribution two different approaches are considered.
The first approach is based on the recorded pressure (or force) drop in time, as it is shown in Fig.~\ref{fig:PressureDrop_Sketch}.
The measured pressure in the hydraulic jack is extracted and monitored in time, with a frequency of $10$~Hz (for tests expected to run longer than several days, frequency of $10$~Hz is used only for the first 24~h, and than reduced to 1~Hz). 
Based on the first obvious drop in the pressure significantly larger than the above mentioned $\pm$2\% hysteresis in the controlled load is defined as the failure time.

\begin{figure*}[h!]
        \begin{center}
        \begin{tabular}{cc}
         \includegraphics[width=0.75\textwidth]{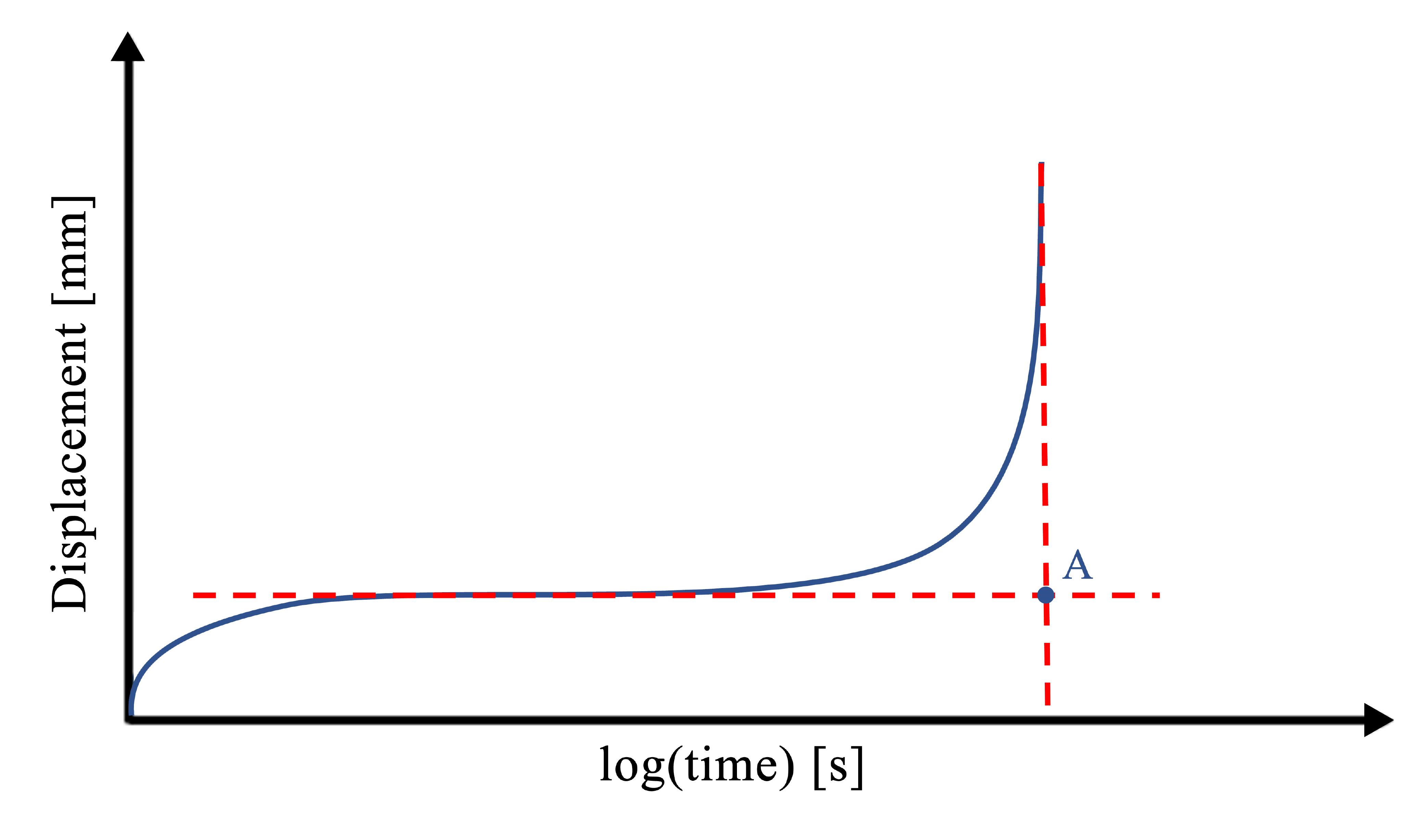}
 		\end{tabular}  
        \caption{Schematic view of the intersection point used to determine time to failure}
         \label{fig:IntersectionPoint}              
        	\end{center}          
     \end{figure*}

The second approach is based on a nominal definition of failure times as intersection point between two linear regression lines in a displacement versus logarithmic time plot, as sketched in Fig.~\ref{fig:IntersectionPoint}.
A linear trend-line is fitted once for the secondary creep stage, for times $0.3t_f\leq t \leq 0.6t_f$, and once for the tertiary creep stage $ 0.9 t_f \leq t \leq t_f$, where $t_f$ the failure time, and $t$ the times used for the regression.
This approach is discussed in detail in \cite{IB_CreepRate}.
A comparison between both approaches revealed minor differences.
In this contribution all reported failure times are defined following the first approach.

\section{Comparison of regression models and recommendations}
\label{Model comparison}

In the following chapter, the data of the systematically carried out experimental investigation serves for the analysis of different regression models with the objective of life-time predictions.
Several approaches as introduced earlier are compared and used to investigate the aforementioned goals.
Finally, a new sigmoid function is found to be a more appropriate choice to relate stress to failure times due to the use of presence of horizontal asymptotes for low and high relative load levels.

\begin{figure}[h!]

\begin{subfigure}[b]{0.5\textwidth}
       \includegraphics[width=1\textwidth]{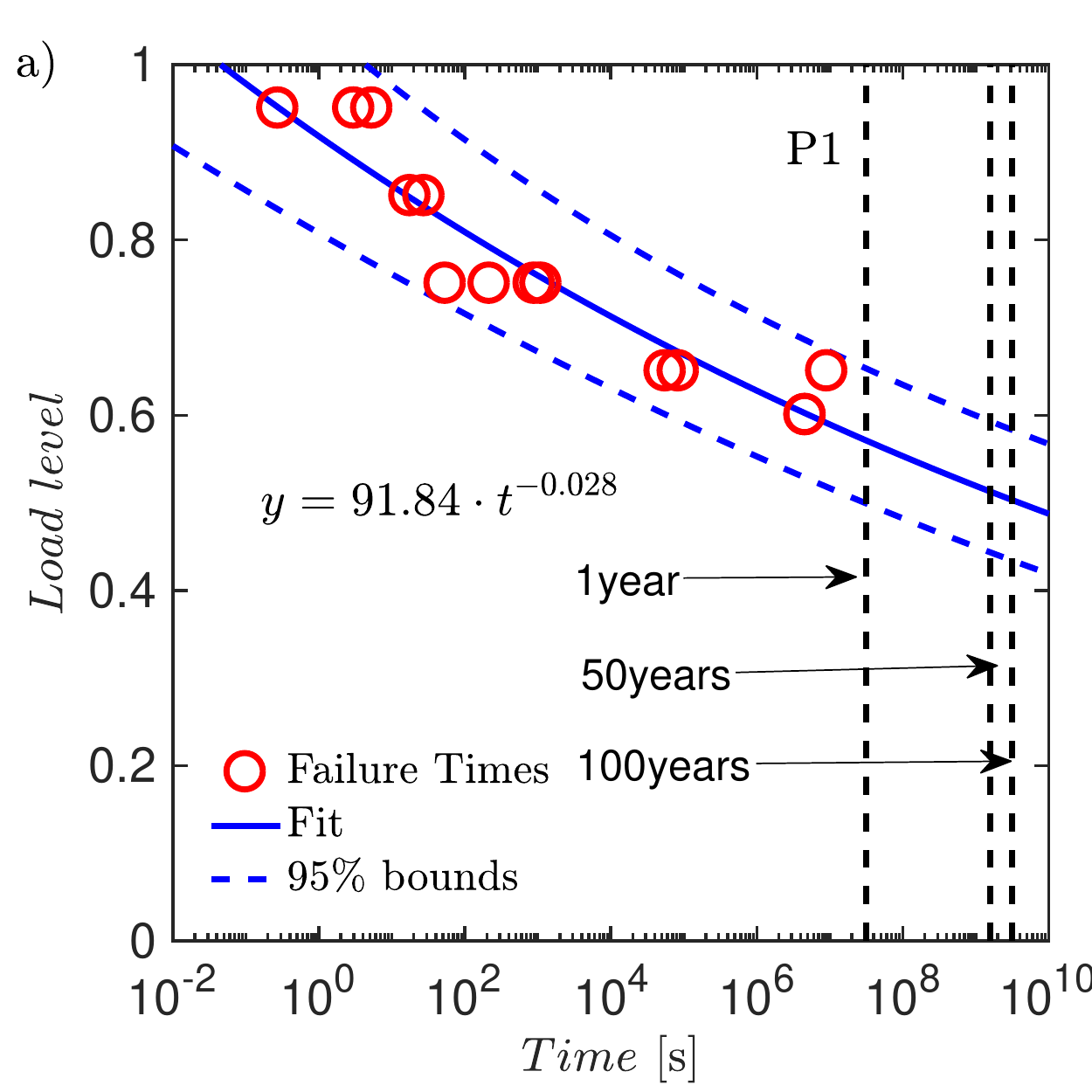}
  \end{subfigure}
   \begin{subfigure}[b]{0.5\textwidth}
      \includegraphics[width=1\textwidth]{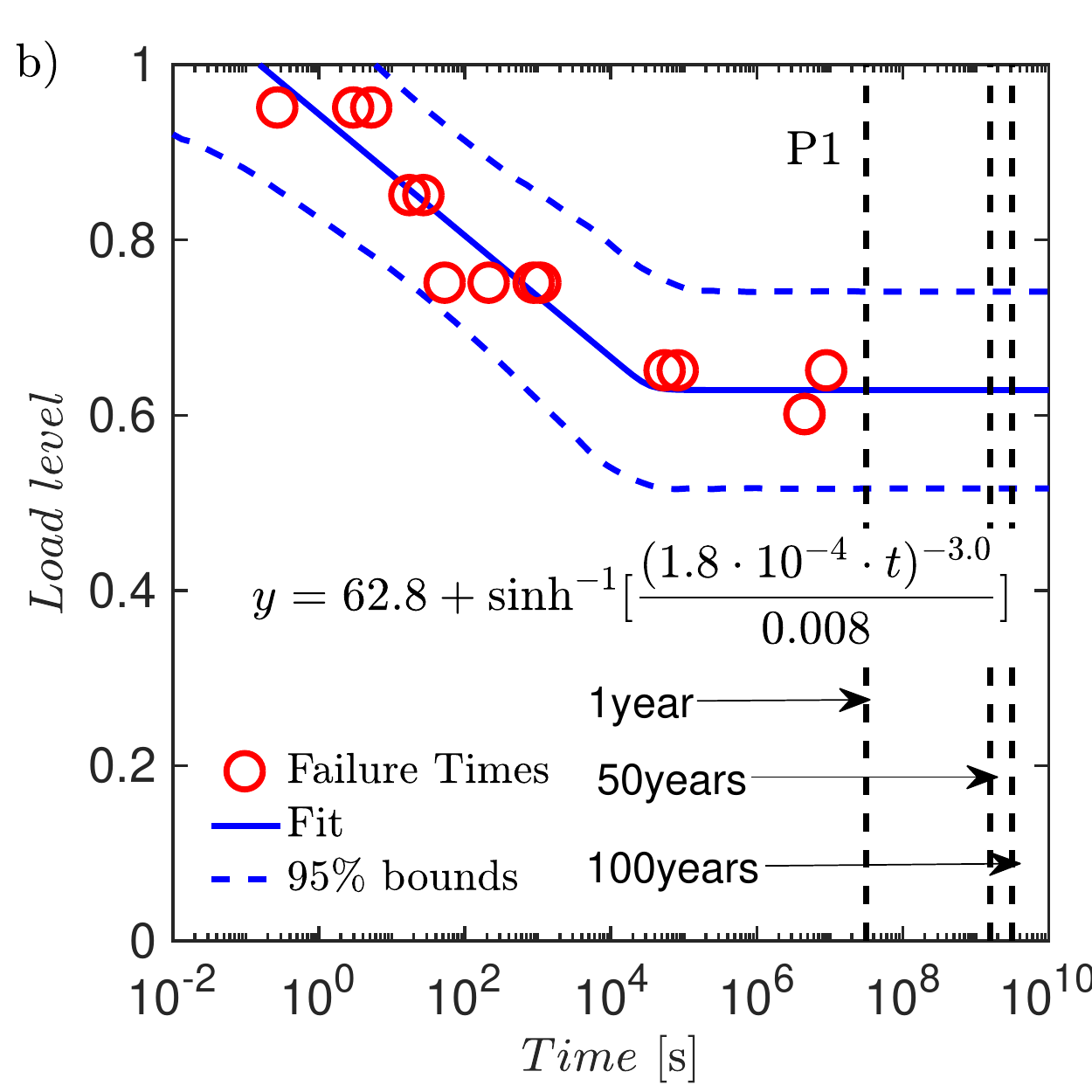}
  \end{subfigure}
  \caption{Product~1: Fit of a power function (a), and fit of rate-theory based function (b)}
  \label{fig:P1_StresvsRatePL}
\end{figure}

\begin{figure}[h!] 
\begin{subfigure}[b]{0.5\textwidth}
       \includegraphics[width=1\textwidth]{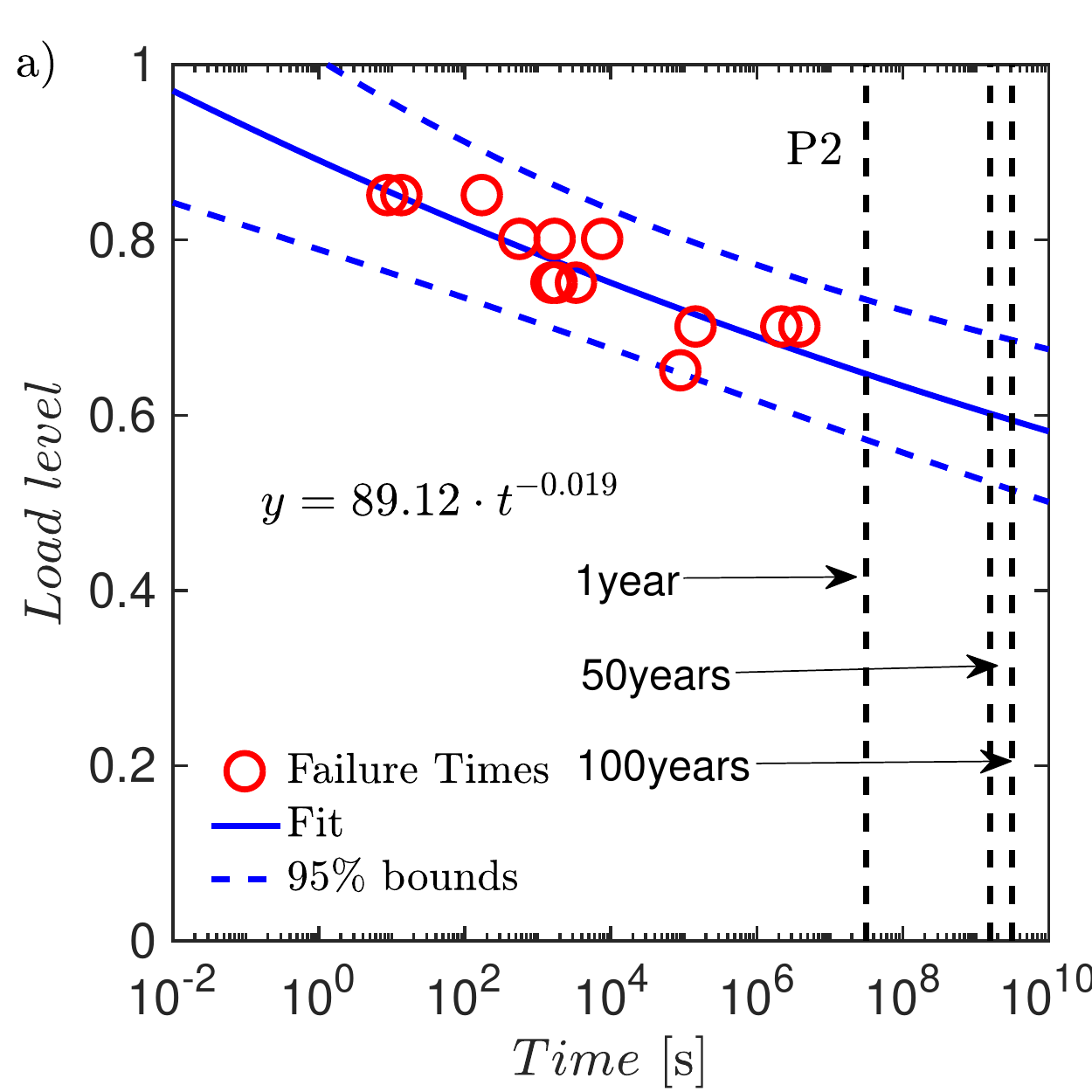}
  \end{subfigure}
   \begin{subfigure}[b]{0.5\textwidth}
      \includegraphics[width=1\textwidth]{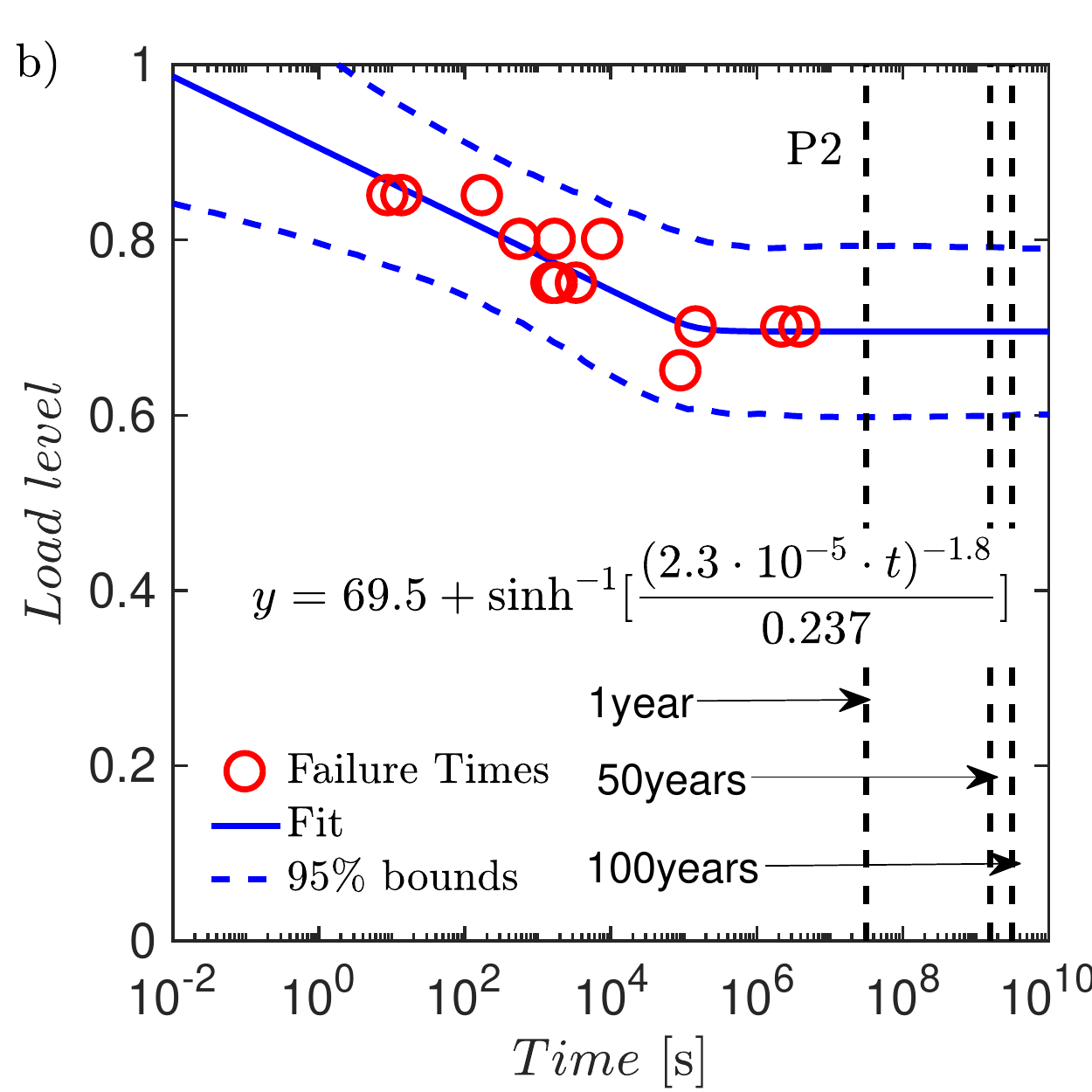}
  \end{subfigure}
  \caption{Product~2: Fit of a power function (a), and fit of rate-theory based function (b)}
  \label{fig:P2_StresvsRatePL}
\end{figure}

The fits of the power function on the experimental data is shown in Fig.~\ref{fig:P1_StresvsRatePL}(a) for Product~1, and in Fig.~\ref{fig:P2_StresvsRatePL}(a) for Product~2.
Fig.~\ref{fig:P1_StresvsRatePL}(b) and Fig.~\ref{fig:P2_StresvsRatePL}(b) show the corresponding fits based on the rate-theory function for Products~1 and 2, respectively.
Both regression models can fit well the experimental data.
The fits agree well with the predicted stress versus time to failure curve as obtained by the creep-rate based approach proposed in \cite{IB_CreepRate}.
The creep-rate based approach reveals a no-failure limit at around 63\% in case of P1, while in case of P2 it occurs to be at around 70\%.

\begin{figure}[h!]

\begin{subfigure}[b]{0.5\textwidth}
       \includegraphics[width=1\textwidth]{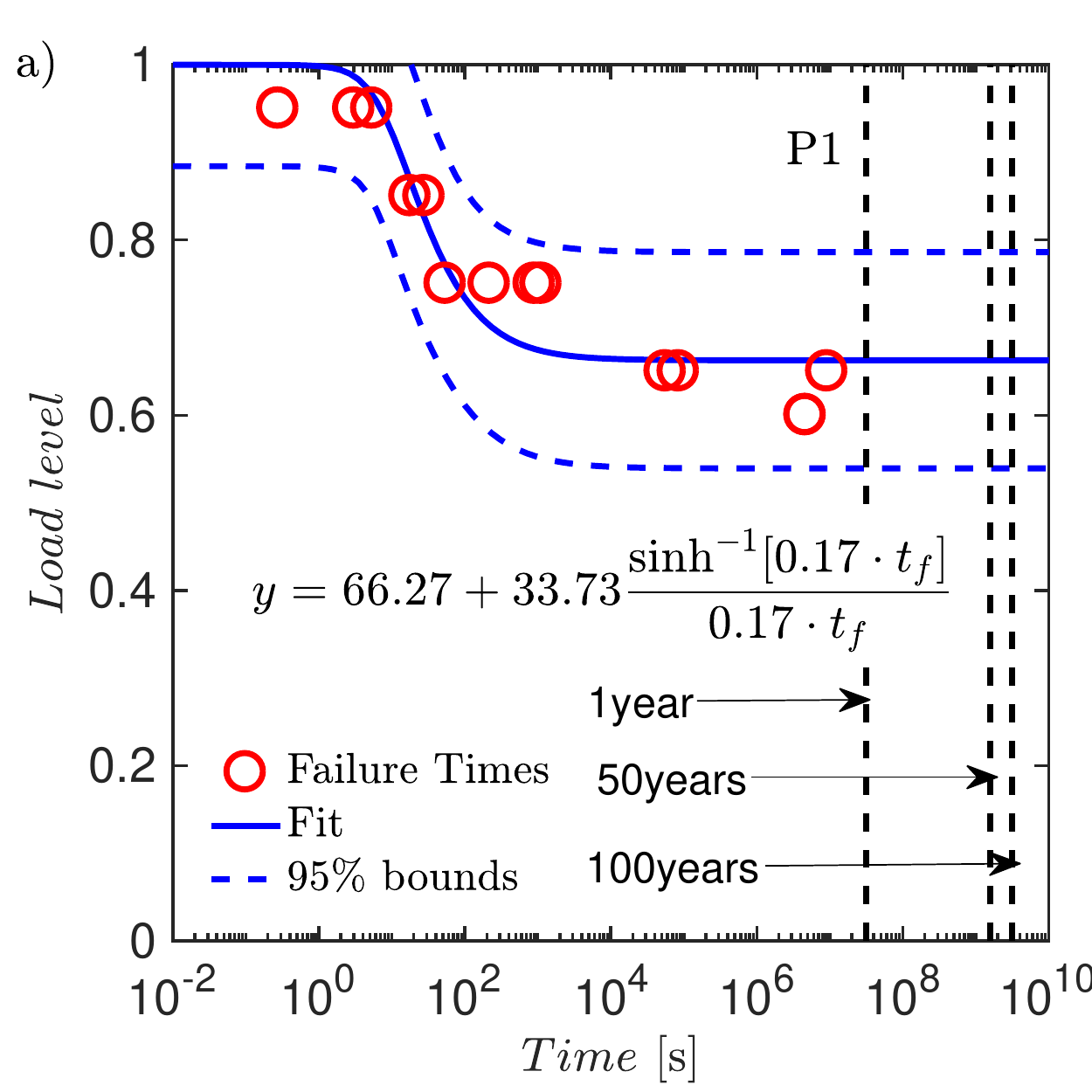}
  \end{subfigure}
   \begin{subfigure}[b]{0.5\textwidth}
      \includegraphics[width=1\textwidth]{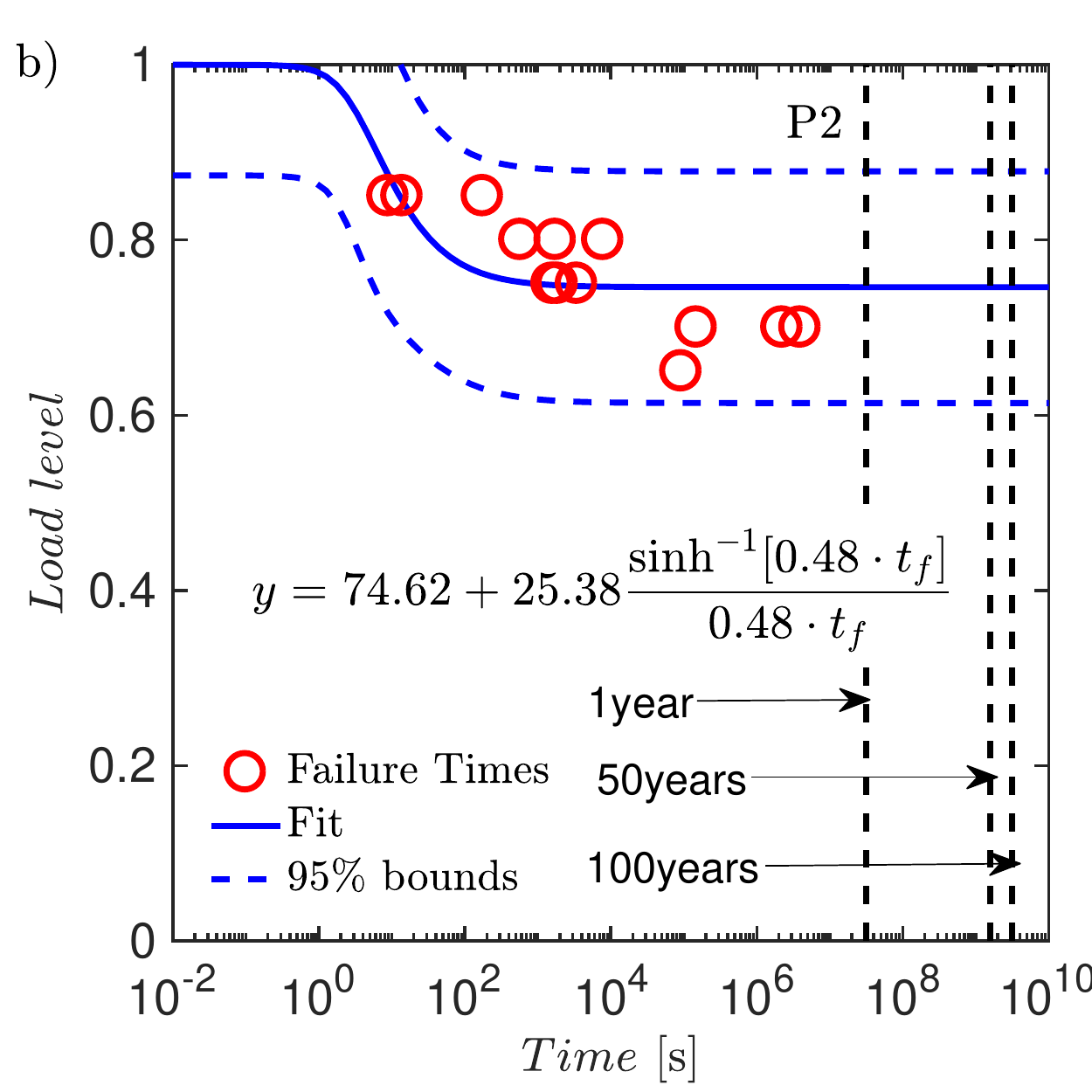}
  \end{subfigure}
 
\caption{Fit of Powell-Eyring equation on stress versus time to failure data of P1 (a) and P2 (b).}
\label{fig:StresvsRateEyring}
\end{figure}

Both models as well as the logarithmic regression model are unable to reproduce (at least approximately) the physical behavior at high relative load levels.
The Powell-Eyring function on the other hand has the advantage of capturing the two asymptotes that are most likely present in this type of tests. 
This model has only three parameters, two representing the asymptotic values and one describing the speed of the transition.
The fit of Eq.~\ref{eq:Eyring} on the experimental data is shown in Fig.~\ref{fig:StresvsRateEyring} for both products.
For all figures, the dashed lines represent the confidence interval of the prediction at level of 95\%.   
If the three parameters are considered independent, a value for $\kappa_{0}$ which is lower than the $100\%$ is obtained. 
This is due to the lack of data at high relative load levels, especially for P2.
Fixing the $\kappa_{0}=100\%$, which is justified by the fact that 100\% load level will lead to theoretically instantaneous failure, further reduces the number of independent variables to two and enforces the correct asymptotic value at high loads.
Finally, the load levels below which (on average) failure will not be observed even at infinite times are determined for the Powell-Eyring model to be $66\%$ for P1 and $75\%$ for P2.


As an alternative, the proposed sigmoid function as introduced in Eq.~\ref{eq:S_ShapeEq} is evaluated.
%
%
The introduced sigmoid function for the stress versus time to failure plot is a non-linear function in logarithmic time.
It is well known that the relationship between stress and TTF is characterized by more than one domain \cite{IB_TertiaryCreep}, which can not be captured by a single linear function.
As discussed for the Powell-Eyring model, the proposed sigmoid function approaches the ``true" asymptotic behavior for high (horizontal asymptote at 100\%) and low stress levels (horizontal asymptote at the threshold to damage development).
In the intermediate (second) domain this function is characterized by an almost linear (in logarithmic time) relationship where typically most of the TTF data points come to rest.
As a consequence, the fit of the regression model is stabilized by adding tests at high load levels. 
These are relatively quick tests albeit complicated by the inherently high scatter that may lead to failure already during loading.
However, if data points are obtained that lie below the linear regression line, i.e. form a concave curve  approaching the asymptotic value 1 (short times), they will improve the long-term extrapolations in the convex part of the sigmoid function (large times) where data points are limited due to the required long test durations.
Naturally, the second asymptote for low load levels can not be estimated if experimental data is restricted to the approximately linear range around the inflection point.

\begin{figure*}[h!]
        \begin{center}
         \includegraphics[width=1\textwidth]{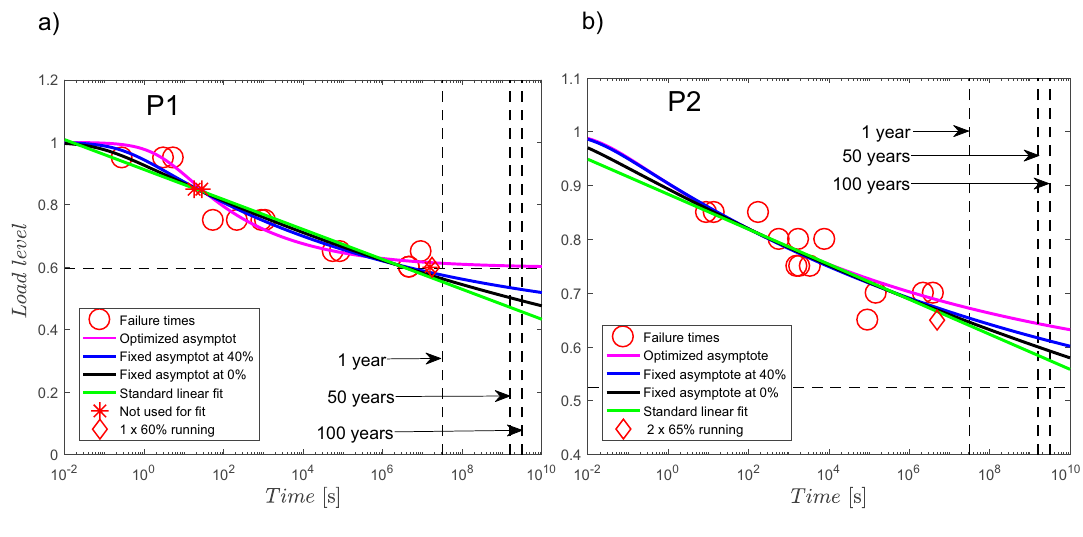} 
        \caption{Comparison of sigmoid function vs. linear regression model for the TTF predictions of a) Product~1 and b) Product~2, respectively}
         \label{fig:Product 1 and 2}              
         	\end{center}          
     \end{figure*}

Fig.~\ref{fig:Product 1 and 2}(a) shows results for Product~1, where the circles represent all tests that failed, and the diamonds the tests that were running at the time the analysis was performed. 
During fitting only data points that actually failed are considered, i.e. results obtained at 95\%, 75\%,65\% and 60\% load levels.
At that time the analysis was performed, two tests at 60\%~ were still running (diamond marker).
Recently,  one of the two running tests at 60\% failed and is only showed in the plot (star marker), but is not used for the fitting analysis.



The data obtained from the sustained load tests at 85\% (marked with stars) are excluded from the fit, and were used to validate the extrapolation method.
As it can be seen, the experimentally obtained failure times are in a very good agreement with the predicted times.
Based on the fit where the $\kappa_{\infty}$ parameter is optimized (parameter $\kappa_{\infty}$ defines asymptotic value), the proposed sigmoid function captures reasonably well the experimental data and flattens out approaching a horizontal asymptote at $\kappa_{\infty}$ = 0.5986 (magenta curve).
That means that fit defines a mean relative load level of 59.9\% as a level below which no failure occurs in a theoretically infinite product life-time.
For a given finite required service life-time the safe load level can be directly obtained by evaluating the function at a given life-time.
Additionally, fits were obtained with two pre-defined asymptotic load levels where the blue and black curve represent fits with fixed asymptotes at 40\% and 0\%, respectively.
The asymptotic load level is the most uncertain model parameter. 
Thus, providing a (conservative) assumption significantly improves the stability of the model and reduces the requirement for very long tests.
The fit with a fixed asymptote at 40\% is chosen as an engineering approach following the common assumption of linear creep (no damage) in a load range of 0-40\% \cite{Eurocode,fib_code_2013}.
The fixed asymptote at 0\% is selected to ensure an infinite time to failure for unloaded systems.
As expected, this approach (black line) is more conservative than the curve with optimized asymptote (magenta line) or fixed asymptote at 40\% (blue line), and still shows higher sustained load strength than the previously promoted approach with a linear model.
Thus, this model can be qualified as overly conservative due to its non-physical properties.
It has to be noted here, that linear trend line (green line) is fitted to all failure times (circles).
According to Cook et al. \cite{Cook_2}, a linear trend line is fitted for the intermediate range (from 85\% to load levels that did not fail in 2000-3000~h) and results in the largest reduction of load levels at 50 years.
Finally, the proposed new approach clearly shows in Fig.~\ref{fig:Product 1 and 2}(a) all limitations of the simple power-law approach (green line), even though it captures most of the practically relevant data (65-80\%).
The proposed function defines an optimized asymptotic relative load level at about 60\% for the investigated product.
The linear fit based including all observed failure times defines a level below which no failure occurs in 50 years at about 45\%. 
If only the intermediate data range (65-80\%) is used for the linear fit, the save 50 year load level would be even more conservative.

Results of the analysis for Product~2 are shown in Fig.~\ref{fig:Product 1 and 2}(b).
As it can be seen, the proposed sigmoid function follows a similar trend compared to the standard linear fit for the investigated range of the load levels due to the nature of the specific product.
Again the optimized fit yields in the least conservative extrapolations followed by fits with fixed asymptotes at 40\% and 0\%. 
The linear fit is the most conservative option resulting in a safe mean relative load level of 58\% for a required life-time of 50 years compared to 64\% for the optimized asymptote.
Clearly the availability of data at higher load levels (above 85\% in this case) combined with the asymptote at 100\% would significantly influence the long-term asymptote. 
Unfortunately, due to the nature of the product type and higher scatter observed at peak tests, all experiments aimed at a load level higher than 85\% failed during loading.
Consequently, most of the obtained data points are concentrated around the inflection point, and as a result all sigmoid fits are similar to the linear fit.
Nevertheless, the proposed function still deviates from the simple linear trend line due to the more physical functional form.

\begin{figure*}[h!]
        \begin{center}

         \includegraphics[width=1\textwidth]{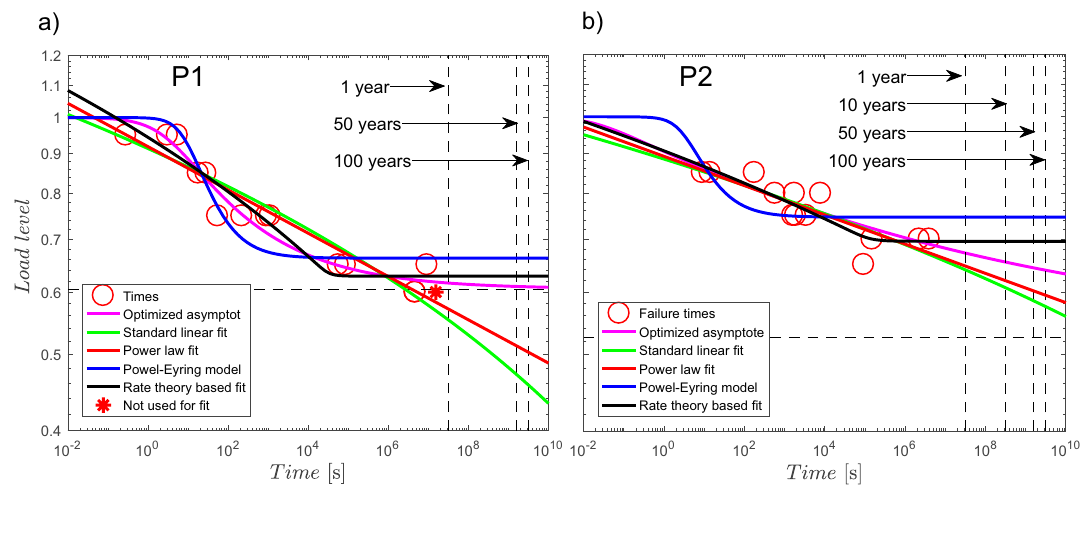}
        \caption{Summary of different fitting models for (a) Product~1, and (b) Product~2}
         \label{fig:Summary}              
         	\end{center}          
     \end{figure*}

A summary of results for different regression models and the available data of Product~1 is shown in Fig.~\ref{fig:Summary}(a). It has to be noted that the summary plot is shown in a log-log plot to stress the differences in functional form while all previous plots were presented in linear load level versus logarithmic time.
Fig.~\ref{fig:Summary}(b) summarizes the corresponding results for Product~2, where clear differences between the different approaches can be seen.


\subsection{Application to available literature data}
\label{Results_Cook_Data}

\begin{figure*}[h!]
        \begin{center}
          \includegraphics[width=1\textwidth]{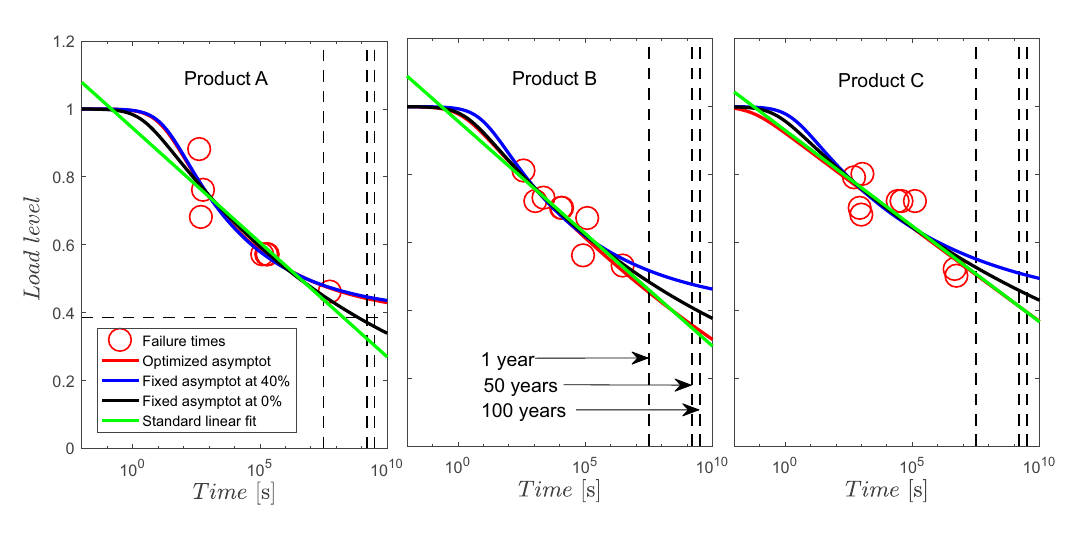} 
        \caption{Application to literature data}
         \label{fig:Subplot_CookData}              
         	\end{center}          
        \end{figure*}

Finally, both the power-law method and the proposed sigmoid function are used to perform the analysis on available literature data \cite{Cook_2,Cook_Report757}.
In general, very limited data sets can be found for the investigated problem, especially reporting all important details in order to perform a systematic analysis.
An experimental study reported by Cook et al. provides several time to failure data sets for different adhesives \cite{Cook_2,Cook_Report757}.
Three of the data sets are included in this contribution, named as A, B and C product.
As reported, different load levels were used for different products, as listed in Tables~\ref{table-Cook_data_Product_A}-\ref{table-Cook_data_Product_C}.
Fig.~\ref{fig:Subplot_CookData} shows the results of the previously introduced analyses to the literature data.

\begin{table}[h!]
\centering
\caption{Reported load levels and failure times for Product A}
\label{table-Cook_data_Product_A}       
\begin{tabular}{lllllllll}
\hline\noalign{\smallskip}
Load level &  88\% & 76\% & 68\% & 57\% & 57\% & 57\% & 57\% & 46\%  \\
\noalign{\smallskip}\hline\noalign{\smallskip}
Failure time [h] & 0.12 & 0.17 & 0.14 & 36 & 52 & 55 & 59 & 16174  \\
\noalign{\smallskip}\hline\noalign{\smallskip}
\end{tabular}
\end{table}


\begin{table}[h!]
\centering
\caption{Reported load levels and failure times for Product B}
\label{table-Cook_data_Product_B}       
\begin{tabular}{lllllllll}
\hline\noalign{\smallskip}
Load level &  81\% & 73\% & 72\% & 70\% & 70\% & 67\% & 56\% & 53\% \\
\noalign{\smallskip}\hline\noalign{\smallskip}
Failure time [h] & 0.11 & 0.67 & 0.32 & 3.29 & 3.6 & 35 & 24 & 862 \\
\noalign{\smallskip}\hline\noalign{\smallskip}
\end{tabular}
\end{table} 
     

\begin{table}[h!]
\centering
\caption{Reported load levels and failure times for Product C}
\label{table-Cook_data_Product_C}       
\begin{tabular}{llllllllll}
\hline\noalign{\smallskip}
Load level &  80\% & 79\% & 72\% & 72\% & 72\% & 70\% & 68\% & 52\% & 50\%  \\
\noalign{\smallskip}\hline\noalign{\smallskip}
Failure time [h] & 0.32 & 0.15 & 11 & 7.76 & 37 & 0.25 & 0.29 & 1347 & 1576 \\
\noalign{\smallskip}\hline\noalign{\smallskip}
\end{tabular}
\end{table} 
     

%
%
     
All results clearly show the deviation from the linear trend line, even though the proposed sigmoid function with positive optimized asymptote can be established only for Product~A while in the other two cases the optimized low load level asymptote would be negative.
The latter is due to the nature of the proposed sigmoid function and a consequence of the data point distribution in the intermediate range around the inflection point rather than in the convex or concave regions.

\subsection{Recommendations for testing}
\label{Recomm-Tests}
     
In general, a long term prediction and extrapolation based on any type of short term tests is a very sensitive and challenging task. 
A reliable and consistently obtained set of experimental data is required.
Specifically for time-to-failure (TTF) test the main possible influence factors are discussed and summarized as follows:
\begin{itemize}
	\item Ensure consistent concrete properties (if possible cast from one batch, cured under well-defined conditions
	\item Minimize aging effects during TTF tests. This can be accomplished my matured concrete members that are at least 90~days old.
	\item Be aware of loading rates effect during the determination of short-term pull-out capacities and the load-application of TTF tests. It clearly shows the effect of the different loading rates and its influence on the pull-out load capacity.
It is crucial to control, or at least to use the same loading rate in short and long term tests in order to avoid any inconsistency defining the load levels for the long term tests with respect to the short term tests.
\item Minimize aging and post-curing effects in the adhesives. Consequently, the time between installation and actual tests needs to be tightly controlled, together with the environmental conditions where the specimens are cured. Ideally, both concrete member and adhesive are stored in a room environmentally control in which also the anchors are later installed and tested. 
\end{itemize} 


\subsection{Recommendations for data analysis}
\label{Recomm-Analysis}
     
Finally, performing the systematic analysis is as important as having a consistent set of experimental data.
Different regression models have been used in the past, and all of them showed some limitations as discussed before.
Depending on the regression model and to ensure the best fitting, it is important to obtain a sufficiently large number of data points with a suitable distribution along the time axis. 
The introduced sigmoid function overcomes most of the limitations that other regression models suffer from while maintaining sufficient conservatisms.
The results of the fitting analysis show the importance of having reliable data at relatively high not only low load levels.
Data points in both regions (deviating from the linear trend in the intermediate domain) are important to stabilize the fit.
The stability of a sigmoid regression model can be increased by prescribing the asymptotic values for high and low loads.
This approach also provides reasonable results in case all data is restricted to the intermediate domain.
A conservative approach can be fixing the asymptotic value at a load level at which definitely no failure should ever occur, e.g. 0\% (= unloaded specimen).
This asymptotic value can also be defined at the commonly accepted region of elastic response such as e.g. 40\% for concrete, which is also suggested by several standards \cite{fib_code_2013,Eurocode} as a load level below which no significant damage occurs in the system.

\section{Summary and conclusion}

This contribution presents a comprehensive experimental campaign including short and long term tests of bonded anchors for two different adhesive products.
The obtained data is used to establish a relationship between load level and related time to failure of bonded anchors under sustained load.
Contrary to the standard approach that uses a liner relationship in semi-logarithmic time only other regression models are explored and investigated with regard to the predictive quality. 
Additionally recommendations for testing and data analysis are derived.
Finally, a new approach based on a sigmoid function is proposed for life-time predictions of bonded anchors subject to sustained loads.

Based on the results of the experimental investigation the following conclusions are drawn:

\begin{enumerate}
			\item Experimental results clearly show the importance of the loading rate effect on strength. 
			Possible loading rate inconsistencies between short and long term tests would lead to inaccurate load level definitions and adversely affect TTF analyses.
		
		\item The commonly assumed linear relationship between stress and logarithmic time to failure leads to overly conservative life-time predictions and can neither represent the behavior of tests at high nor at low load levels.
		A conservative choice of $\kappa_{\infty}$ simplifies the analysis and ensures sufficiently but overly conservative predictions.

		\item A sigmoid regression model with horizontal asymptotes at high and low loads can represent that actual shape of TTF data better without increasing the number of regression parameters. 
		
		\item Short-term TTF data at very high load levels (in the concave region) can help to stabilized long-term predictions at low load levels (convex region). 
		Data restricted to the intermediate linear domain around the inflection point contains only limited value for TTF extrapolations. 

\end{enumerate}

\section*{Acknowledgements}

The financial support by the Austrian Federal Ministry for Digital and Economic Affairs and the National Foundation for Research, Technology and Development is gratefully acknowledged.

\section*{References}


\end{document}